\documentclass[journal]{new-aiaa} 
\usepackage[utf8]{inputenc}

\usepackage{placeins}
\usepackage{grffile}
\usepackage{supertabular}
\usepackage{multicol}
\usepackage{multirow}
\usepackage{graphicx}
\usepackage{amsmath}
\usepackage[version=4]{mhchem}
\usepackage[per-mode=symbol]{siunitx}
\usepackage{longtable,tabularx}
\usepackage{caption}
\usepackage{subcaption}
\usepackage{booktabs}
\usepackage{floatrow}
\usepackage{tikz}
\usepackage{bm}
\usetikzlibrary{arrows.meta,arrows}
\newcolumntype{x}[1]{>{\centering\arraybackslash\hspace{0pt}}p{#1}}
\DeclareSIUnit{\million}{\text{M}}
\title{Parameter Influence on Porous Bleed Performance for Supersonic Turbulent Flows}

\author{Julian Giehler\footnote{Ph.D. Student, Aerodynamics, Aeroelasticity, Acoustics Department, julian.giehler@onera.fr}}
\author{Pierre Grenson\footnote{Research Scientist, Aerodynamics, Aeroelasticity, Acoustics Department, pierre.grenson@onera.fr}}
\author{Reynald Bur\footnote{Senior Research Scientist, Aerodynamics, Aeroelasticity, Acoustics Department, reynald.bur@onera.fr}}
\affil{DAAA, ONERA, Universit\'e Paris Saclay, 92190 Meudon, France}

\begin{document}
\DeclareGraphicsExtensions{.pdf,.png,.jpg,}

\maketitle
\begin{abstract}
    Porous bleed systems are a common technique to control shock-boundary layer interactions and/or supersonic boundary layers. However, the influence of various design parameters is still unknown. Even though bleed models are required to minimize the costs of the design process, they often do not include parameter effects. In the present study, the effect of the plate length, the hole diameter, the porosity, the thickness-to-diameter ratio, and the stagger angle are investigated by means of three-dimensional RANS simulations. The bleed efficiency and the effectiveness in thinning a Mach $M=1.6$ turbulent boundary layer are determined. The findings show a crucial influence of the hole diameter on both efficiency and effectiveness of the porous bleed. Similar findings are made for the porosity and stagger angle but with a smaller significance. Thickness-to-diameter ratio and plate length are shown to mainly affect the bleed efficiency.
\end{abstract}

\section{Nomenclature}

\noindent\textit{Latin symbols}

\begin{supertabular}{@{}l @{\quad=\quad} l@{}}
    $A$  & Area \\
    $A_o$  & (Circular) extraction area \\
    $A_{pl,ex}$  & Plenum exit throat area \\
    $D$  & Hole diameter \\
    $H$  & Boundary layer shape factor \\
    $L$  & Length \\
    $M$  & Mach number \\
    $\dot{m}$  & Mass flow rate \\
    $p$  & Static pressure \\
    $Q_{sonic,w}$  & Surface sonic flow coefficient \\
    $R$  & Specific gas constant (air)\\
    $Re$  & Reynolds number\\
    $T$  & Static temperature \\
    $T/D$  & Thickness-to-diameter ratio \\
    $TR$  & Throat ratio \\
    $v$  & Transpiration velocity \\
    $x,y,z$  & Cartesian coordinates in streamwise, wall-normal, and spanwise direction \\
    $\hat{x}$, $\check{x}$  & Streamwise position from start/end of bleed region \\
\end{supertabular}\\

\noindent\textit{Greek symbols}

\begin{supertabular}{@{}l @{\quad=\quad} l@{}}
    $\alpha$  & Angle of attack \\
    $\beta$  & Stagger angle \\
    $\delta$ & Boundary layer thickness \\
    $\delta_1$  & Displacement thickness \\
    $\varepsilon_{\tau}$  & Rise in wall shear stress \\
    $\gamma$  & Heat capacity ratio \\
    $\phi$  & Porosity level \\
    $\rho$  & Density \\
    $\tau$  & Shear stress \\
\end{supertabular}\\

\noindent\textit{Subscripts}

\begin{supertabular}{@{}l @{\quad=\quad} l@{}}
    $bl$  & Bleed\\
    $c$  & Compressible\\
    $h$  & Hole\\
    $pl$  & Plenum\\
    $sonic$  & Sonic \\
    $t$  & Total \\
    $w$  & External wall\\
    $\infty$  & Free-stream\\
    $99$  & 99 \% boundary layer thickness \\
\end{supertabular}

\section{Introduction}
\begin{figure}[ht!]
    \centering
    \includegraphics[]{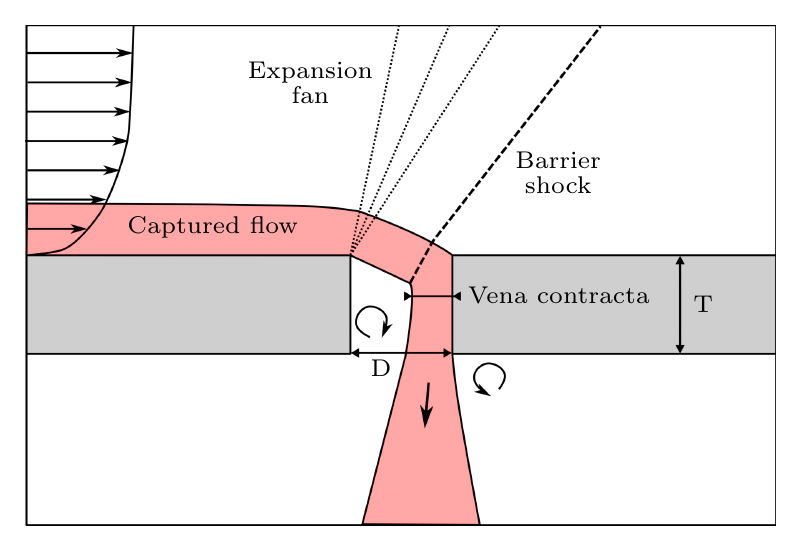}
    \caption{Schematic of the flow through a porous plate in supersonic flow}
    \label{fig:schematic}
\end{figure}

\lettrine{T}{he} application of a porous bleed is a proven technology to mitigate the boundary layer separation caused by shock-boundary layer interactions in various applications, such as supersonic air inlets. Typical bleed systems consist of a vast number of small holes which cover the area around or upstream of the shock foot. The principle is to remove the low-momentum air near the wall and, as a result, to generate a boundary layer profile that is less susceptible to separation for positive pressure gradients (see Fig.~\ref{fig:schematic}). Crucial for the system's success is a sufficient ratio between the wall pressure and the pressure in the bleed plenum, which is located below the holes. Up to now, it remains challenging to predict the required pressure to remove the desired mass and also to estimate the effect of the bleed system on the boundary layer and, consequently, on the shock-boundary layer interaction.

In the past, several researchers investigated the effect of porous bleed experimentally. In contrast to real applications, the porous bleed regions in experiments were scaled in size, and only a small number of rows~\cite{Willis1995a,Willis1995b,Willis1996,Akatsuka2006,Oorebeek2012,Oorebeek2013,Oorebeek2015}, or even a single hole~\cite{Bodner1996,Davis2012,Eichorn2013}, has been investigated. Moreover, the porous bleed has been globally analyzed, which means that only a little information about the flow in the individual holes exists~\cite{Hingst1983,Benhachmi1989}. As a result, the longitudinal evolution of the local mass removal has not been studied, and all bleed holes have been supposed to work equally. Moreover, the knowledge about the influence of the bleed parameters (e.g., hole diameter, hole thickness-to-diameter ratio, porosity, etc.) is limited. Only a few studies focused on various plate geometries~\cite{Willis1995b,Willis1996,Davis1997,Akatsuka2006,Davis2012,Eichorn2013}.

In contrast to experiments, numerical investigations make extracting the mass flow rates for individual holes feasible. Nevertheless, recent numerical studies~\cite{Slater2012,Teh2013,Oorebeek2015,Choe2018,Zhang2020,Akar2020} lack to show and analyze the local mass flow rates. Also, the change of the flow field in the holes along the porous bleed and around the hole has not been analyzed. Similar to the experiments, the bleed region has been observed globally without focusing on the interaction between the holes. Differences in the flow field downstream of the bleed region were out of focus, even though effective control is the primary goal of applying porous bleed systems.

The need for more information about the local flow behavior complicates finding a suitable model to capture the flow physics. Bleed models are unavoidable in the industrial context since the accurate meshing of all holes makes even Reynolds-Averaged Navier–Stokes (RANS) simulations very cost- and time-consuming. Bleed models can be applied as local boundary conditions without refining the mesh. Nevertheless, common bleed models~\cite{Doerffer2000,Slater2012} also neglect the influence of bleed parameters and the hole's location. To the author's knowledge, only a few models consider bleed parameters or local porosity variations~\cite{Harloff1996,Akatsuka2006,Choe2018,Grzelak2021, Giehler2022c}. More precise data are necessary to understand the working principle of bleed systems and create a more accurate bleed model valid for an extensive range of parameters.

In this investigation, a large dataset of bleed configurations is presented. The bleed parameters porosity, hole diameter, hole thickness-to-diameter ratio, stagger angle, and plate length are evaluated, and their influence on both the \textit{efficiency} and \textit{effectiveness} is illustrated. The term \textit{efficiency} relates to the ability to remove a large amount of air by applying a small pressure difference between the external wall and the cavity. On the contrary, the ability to thin the boundary layer is stated as the \textit{effectiveness}. As found in this study, \textit{efficiency} and \textit{effectiveness} are unrelated and must be characterized separately. In contrast to previous studies, all physical quantities are locally computed, allowing us to improve the understanding of the operation of a porous bleed.

The paper is organized as follows: In Sec.~\ref{sec:problem}, we introduce the physical domains to investigate the influence of the parameters. The numerical methodology is then explained in Sec.~\ref{sec:method}, and the validation of the numerical setup is presented in Sec.~\ref{sec:validation}. In the following section, the results of the parametric study are discussed. Our conclusions are drawn in the final section.


\section{Description of the problem} \label{sec:problem}

The investigated phenomenon in this study is the supersonic turbulent boundary layer bleeding for a Mach number of $M=1.6$. The porous bleed system is installed on a flat plate, where it thins a turbulent boundary layer like a bleed system installed upstream of an incident shock. Thus, the influence of the bleed parameters for supersonic conditions is investigated.

\begin{figure}[ht!]
    \centering
    \begin{subfigure}[t!]{0.60\textwidth}
        \centering
        \includegraphics[trim={27mm 31mm 0mm 46mm},clip]{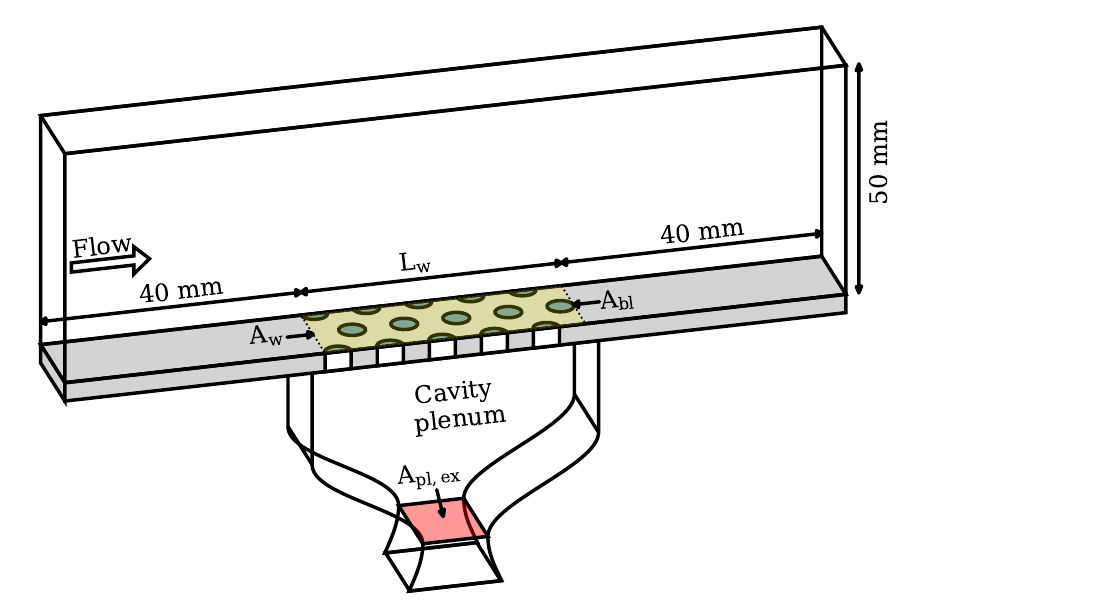}
        \caption{Domain for the numerical investigation}
        \label{fig:domainFlatPlateIso}
    \end{subfigure}
    \begin{subfigure}[t!]{0.28\textwidth}
        \centering
        \includegraphics[trim={3mm 2.3mm 2mm 2mm},clip]{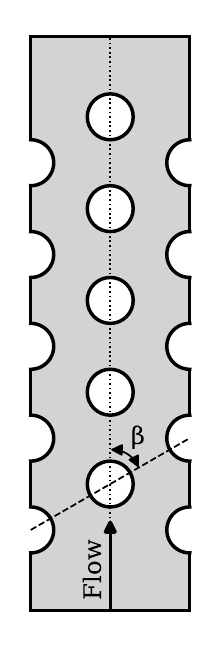}
        \caption{Schematic of the hole pattern}
        \label{fig:domainFlatPlateTop}
    \end{subfigure}
    \caption{Computational domain for the supersonic boundary layer bleeding}
    \label{fig:domainFlatPlate}
\end{figure}

Fig.~\ref{fig:domainFlatPlate} illustrates the flow domain. The investigated bleed region starts with an offset of \SI{40}{\mm} from the inlet, and its length $L_w$ is up to \SI{80}{\mm}. In the spanwise direction, three bleed holes are included in the domain, cutting the first and the third one in the center, as shown in Fig.~\ref{fig:domainFlatPlateIso}. Thus, the width of the porous plate corresponds to twice the spanwise distance between the hole columns. The staggering of the hole columns (see Fig.~\ref{fig:domainFlatPlateTop}) is altered in the steps $\beta=$ \SIlist{30;45;60;90}{\degree} with $\beta=$ \SI{90}{\degree} being an orthogonal hole pattern.

The inflow properties are shown in Tab.~\ref{tab:SimProperties}, based on the flow conditions in the ONERA S8Ch wind tunnel in Meudon, where experiments are planned. All quantities are extracted \SI{10}{\mm} ($\approx 2.5 \delta_{99}$) upstream of the first bleed hole, where the flow is unaffected by the suction independently of the plate geometry.

\begin{table}[h!]
    \centering
    \begin{tabularx}{\textwidth}{x{0.08\textwidth}x{0.09\textwidth}x{0.08\textwidth}x{0.12\textwidth}x{0.08\textwidth}x{0.08\textwidth}x{0.08\textwidth}x{0.08\textwidth}x{0.08\textwidth}}
         \toprule
         $M$ &$p_t$ &$T_t$ &$Re_x$ & $Re_{\delta_2}$ & $\delta_{99}$ & $\delta_{1,c}$ & $\delta_1$ & $H$ \\
         \midrule
         \num{1.6} &\SI{93000}{\Pa} &\SI{300}{\K} &\SI{12.7d6}{\per\m} & \num{5415} & \SI{4.16}{\mm} & \SI{.87}{\mm} & \SI{.57}{\mm} & \num{1.36} \\
         \bottomrule
    \end{tabularx}
    \caption{Physical properties of the flow \SI{10}{\mm} upstream of the first bleed hole}
    \label{tab:SimProperties}
\end{table}

The base area of the plenum is equal to the bleed region's area, with its depth being \SI{20}{\mm}. Below the plenum, a choked convergent-divergent nozzle is added to fix the plenum exit mass flow rate. The throat ratio

\begin{equation}
    TR = \frac{A_{pl,ex}}{A_{bl}} = \frac{A_{pl,ex}}{\phi A_{w}} \label{eq:throatRatio}
\end{equation}
is introduced to compare different bleed parameters for identical conditions. It is the ratio of the areas of the exit nozzle throat $A_{pl,ex}$ and the bleed region $A_{bl}$, which is, in turn, a function of the porosity level $\phi$ and the plate area $A_{w}$.



\section{Numerical methodology} \label{sec:method}

The parametric study of the porous plate geometry is performed using steady-state numerical simulations. The used setup is described in this section.

\subsection{Geometry and mesh}

\begin{figure}[b!]
    \begin{floatrow}
        \ffigbox[\FBwidth]{\includegraphics[width=.525\textwidth]{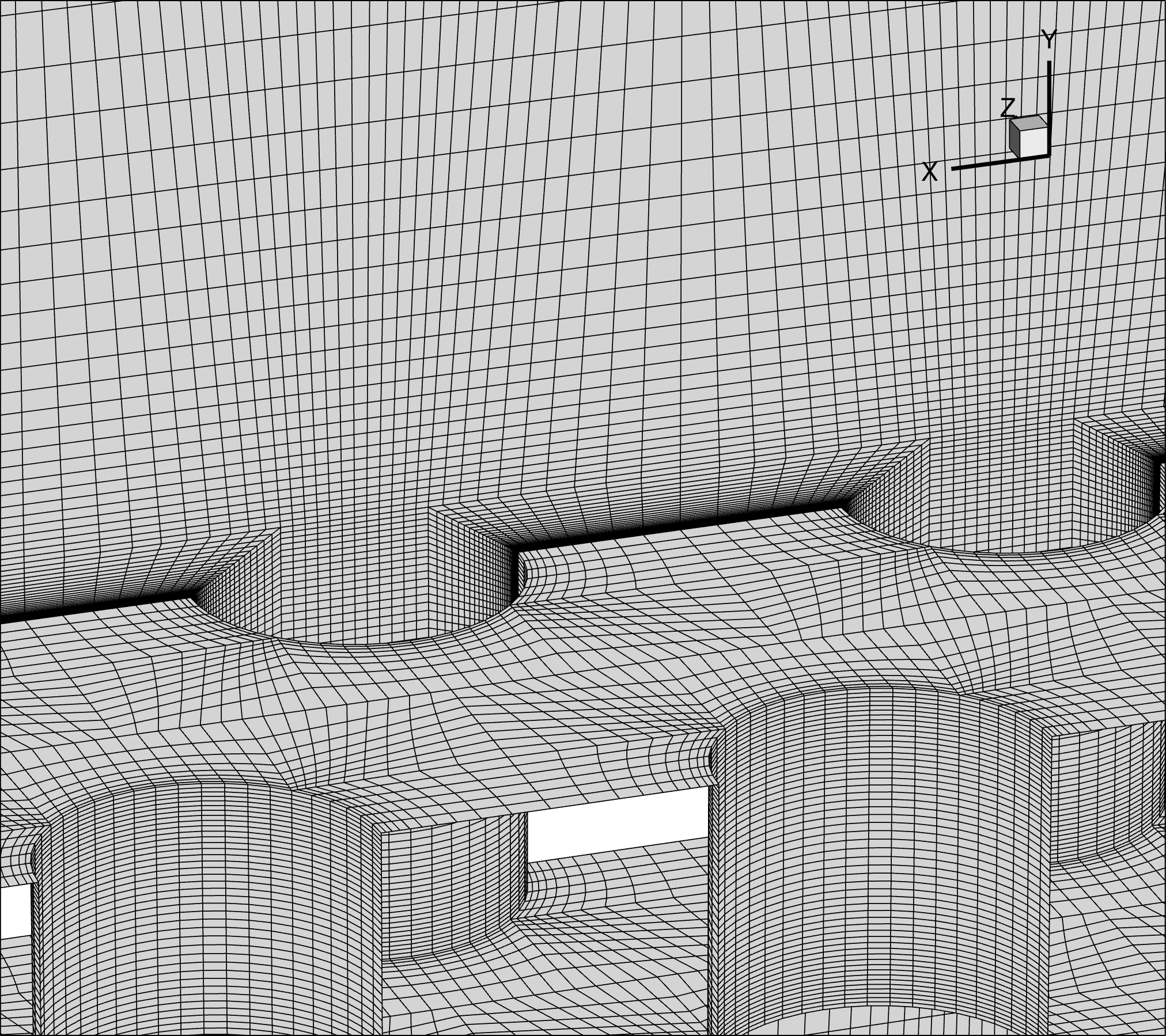}}
                          {\caption{View on mesh around the holes}\label{fig:isoview}}
        \hspace{5mm}
        \ffigbox[\FBwidth]{\includegraphics[trim={26mm 4.7mm 9mm 19mm},clip]{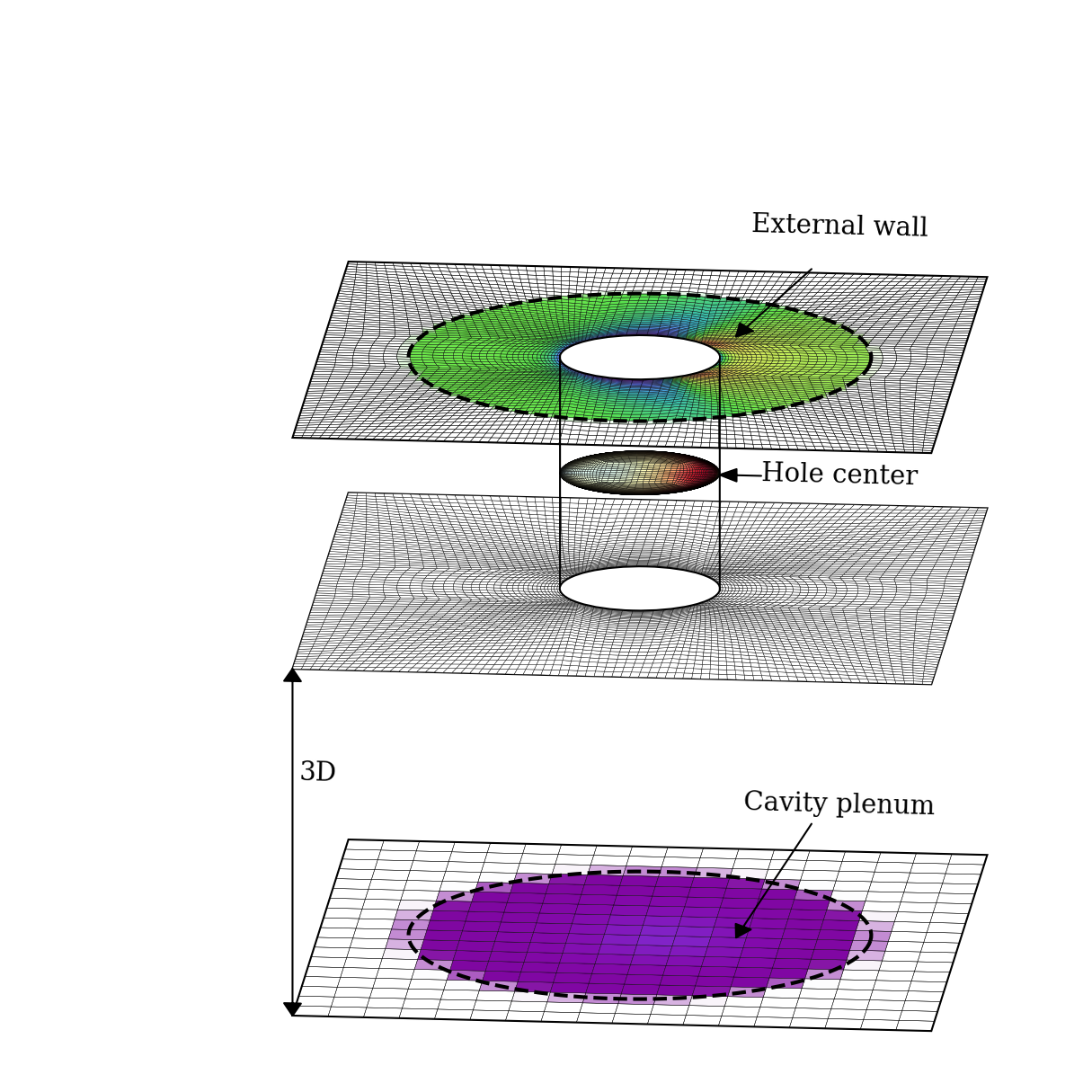}}
                          {\caption{Positions for the local extractions}\label{fig:extract}}
    \end{floatrow}
\end{figure}

The structured mesh is generated using the in-house pre-processing tool and mesh-generator \textbf{Cassiopee}~\cite{Benoit2015}. A fully parameterized mesh allows the variation of the bleed parameters. Therefore, each hole is modeled out of five blocks using a butterfly mesh. The four other blocks are part of a C-grid, including the wall boundary layer on the plate and plenum sides (see Fig.~\ref{fig:isoview}). For a proper resolution of the boundary layer, the minimum wall-normal cell size is $y^+ \approx 1$ (\SI{.2e-3}{\mm}). Inside the holes, the wall is equally meshed. The cell-to-cell growth ratio is \num{1.1}, the maximum value to accurately predict the local mass flow rate, thanks to a preliminary mesh sensitivity study. The total number of cells is between \SI{1.5}{\million} and \SI{22.2}{\million} for a plate length $L_w=$ \SI{40}{\mm}.

\subsection{Flow solver}

The compressible Navier-Stokes equations are numerically solved using the ONERA-Safran finite-volume solver \textbf{elsA}~\cite{Cambier2013}. The Spalart-Allmaras turbulence model with quadratic constitutive relation~\cite{Spalart2000} is applied. The second-order Roe upwind scheme together with the minmod limiter and the Harten entropic correction are used for the spatial derivative with a backward-Euler implicit local time-stepping scheme. A supersonic turbulent boundary layer profile is set as the inlet. The plenum exit and the main outlet are supersonic outlets, while the far-field boundary condition is applied at the top of the domain. Top and bottom walls, and all the walls around and inside the holes are no-slip walls, while a slip-wall is applied at the plenum side walls to reduce the mesh size inside the cavity. The domain is limited using a periodic boundary condition on the front and back.

The bleed flow is fixed by setting a throat ratio. Thus, the plenum-to-external-wall pressure ratio is similar for all plates if the throat ratio remains constant, which allows the evaluation of the bleed efficiency for different configurations. In contrast, a constant bleed mass flow rate is required to quantify the effectiveness. In this case, the throat ratio is automatically adjusted to obtain equal bleed mass flow rates for all hole configurations.

\subsection{Extraction and evaluation of the bleed efficiency and effectiveness}

The efficiency of porous bleed systems is typically quantified by the sonic flow coefficient. In this study, the scaling of~\citet{Slater2012} is used to compute the surface sonic flow coefficient

\begin{equation}
    Q_{sonic,w} = \frac{\dot{m}_{bl}}{\dot{m}_{sonic,w}},
    \label{eq:qsonicw}
\end{equation}
where the surface sonic mass flow rate $\dot{m}_{sonic,w}$ normalizes the bleed mass flow rate $\dot{m}_{bl}$. The surface sonic mass flow rate can be computed using the static flow quantities at the wall as follows:

\begin{equation}
    \dot{m}_{sonic,w} =  p_w A_{bl}\left(\frac{\gamma}{R T_w}\right)^{1/2}\left(\frac{\gamma+1}{2}\right)^{-\frac{\gamma+1}{2(\gamma-1)}}
    \label{eq:msonic}
\end{equation}

With the aim to analyze the local mass flow rate for each hole, the local extraction of the hole mass flow rate $\dot{m}_{bl,i}$ and the local wall pressure is required. In the current study, the mass flow rate is extracted locally in each hole instead of using the bleed mass flow rate extracted at the plenum exit. Therefore the flow momentum in the plate's normal direction is integrated

\begin{equation}
    \dot{m}_{bl,i} = \int_S \rho v dS
\end{equation}
at the hole center corresponding to half the plate thickness, as illustrated in Fig.~\ref{fig:extract}. The (global) bleed mass flow rate can be easily computed by summing all hole mass flow rates:

\begin{equation}
    \dot{m}_{bl} = \sum_{i=1}^N \dot{m}_{bl,i}
\end{equation}

The static wall pressure and temperature cannot be extracted at the hole position since its value is disturbed by the local velocity caused by the flow into the holes. Therefore, a circular patch around the hole is defined with its size $A_{o}$ fitting the porosity $\phi$ concerning the hole area $A_h$:

\begin{equation}
    \phi = \frac{A_{bl,i}}{A_{o}}
\end{equation}

The external wall quantities are then extracted by averaging their values above the patch size (see Fig.~\ref{fig:extract}). Cells that partially lie within the circular patch are weighted using the areal fraction inside the circle. A similar method is applied to the plenum pressure, where the patch is equal in size but placed three hole diameters below the wall instead of using the wall pressure, which can be highly affected by the under-expanded jet at the hole exit. Higher patch distances lead to local information losses as the plenum pressure becomes more uniform with further distance from the holes.

The bleed effectiveness describes the ability of the bleed system to control the flow. In this investigation, the aim of the porous bleed is boundary layer thinning. Therefore, characteristic quantities of the boundary layer are required. The logical way to quantify the boundary layer thinning is to extract the boundary layer thickness up- and downstream of the bleed region and their comparison. However, this is challenging even in this relatively simple case of a flat plate as shock waves and expansion fans induced by the suction holes profoundly modify the boundary layer downstream of the region. Depending on the used definition of the boundary layer thickness, the obtained results are not comparable. Moreover, the shape of the boundary layer profile is not considered, while a full boundary layer profile is desired.

Another quantity that allows conclusions about the shape of the boundary layer profile is the wall shear stress or friction coefficient. The fuller and/or thinner the boundary layer, the higher the wall shear stress. Thus, comparing the wall shear stress up- and downstream of the bleed regions enables the evaluation of the effectiveness. Therefore, the wall shear stress along the span is extracted on a line \SI{100}{\mm} upstream of the leading edge of the first hole and \SI{100}{\mm} downstream of the rear edge of the last hole. Using these quantities, we define the rise in the wall shear stress as

\begin{equation}
    \varepsilon_{\tau} = \frac{\tau_{w,d}-\tau_{w,u}}{\tau_{w,u}}
\end{equation}
quantifying the relative increase of the wall shear stress along the bleed region. However, it is essential to note that this value is case-sensitive and dependent on the inflow conditions, which are kept constant in this study.

Moreover, the bleed mass flow rate must be constant relative to the inflow to evaluate the effectiveness. Hence, the displacement mass flow rate

\begin{equation}
    \dot{m}_{\delta_{1,c}} = \int \delta_{1,c} \rho_{\infty} u_{\infty} dz,
\end{equation}
which describes the theoretical missing mass flow rate required to obtain an inviscid flow, is computed. The bleed mass flow rate is set to remain a constant ratio of these two values.


\section{Validation with experimental data} \label{sec:validation}

The validation of the numerical setup, including the parametric mesh, is conducted based on two experimental datasets. In the first step, a single-hole approach is used to compare the simulation with the data obtained by~\citet{Eichorn2013}, where three different plate geometries are simulated. Moreover, the experimental data of the investigation of~\citet{Willis1995b} for the C1 plate are used to validate the multi-hole setup.

The inlet profiles are reproduced for both setups by performing flat plate simulations with identical total conditions. The inlet profiles are extracted at the position where the boundary layer thickness $\delta_{99}$ is equal to the experimental data. Furthermore, the external wall pressure is not locally extracted as in the following sections but computed out of total pressure and Mach number by applying the isentropic relation equal to the experiments.

\subsection{Single-hole validation} \label{sec:Eichorn}

\citet{Eichorn2013} measured the mass flow rate for \num{21} plates with different inclination angles, thickness-to-diameter ratios, and hole diameters for Mach numbers $M=$ \SIlist{1.33;1.62;1.97;2.46;2.96}{}. In this study, we only focus on the \SI{90}{\degree}-holes and the measurements for a Mach number of $M=1.62$. Thus, three plates with varying hole diameters and thickness-to-diameter ratios are selected for the validation; their geometrical parameters are shown in Tab.~\ref{tab:platesEichorn}. It is crucial to know that the number of holes $N_z$ on the plate is larger for the smallest hole, which was required for the mass flow rate measurements to obtain an almost constant bleed area.

\begin{table}[h!]
    \centering
    \begin{tabularx}{0.50\textwidth}{x{0.10\textwidth}x{0.10\textwidth}x{0.10\textwidth}x{0.10\textwidth}}
         \toprule
         Plate &$D$ &$T/D$ &$N_z$ \\
         \midrule
         101  &\SI{6.350}{\mm} &2.000 &1 \\
         102  &\SI{6.350}{\mm} &1.125 &1 \\
         1501 &\SI{0.794}{\mm} &2.000 &15 \\
         \bottomrule
    \end{tabularx}
    \caption{Simulated plates from~\citet{Eichorn2013}}
    \label{tab:platesEichorn}
\end{table}

\begin{figure}[b!]
    \centering
    \includegraphics[]{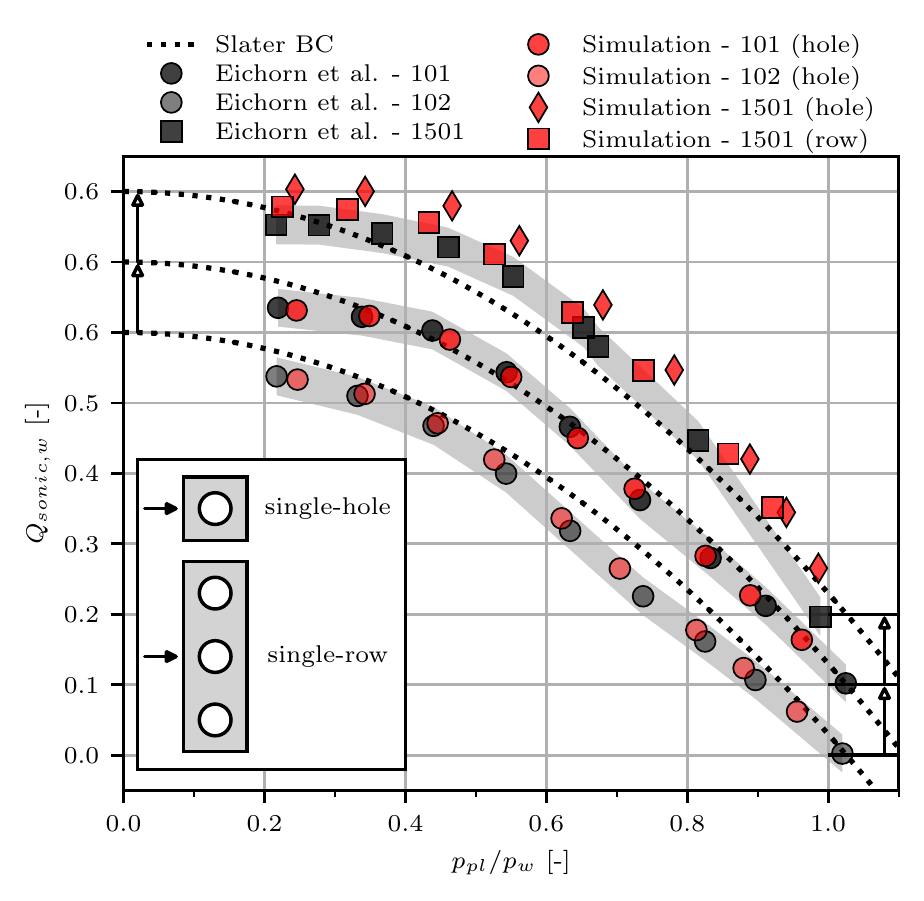}
    \caption{Experimentally and numerically determined surface sonic flow coefficient for the plates of~\citet{Eichorn2013} compared to the regression of~\citet{Slater2012}; plates \SIlist{101;1501}{} are shifted upwards to increase the readability of the graph; gray area illustrates \SI{5}{\%} range around the experimental data}
    \label{fig:validationEichorn}
\end{figure}

The results of the numerical investigation are compared to the data of~\citet{Eichorn2013} and the regression of~\citet{Slater2012} in Fig.~\ref{fig:validationEichorn}. The numerical simulations have a good fit with the experiments, especially for the plates \SIlist{101;102}{}, where the simulations mirror almost perfectly the trend of the experiments.

For plate 1501, two different numerical setups are used: a single-hole approach and a single-row approach, which is equal to the multi-hole approach but with one hole in the streamwise direction only and thus similar to the experiments. \citet{Eichorn2013} stated that no spanwise interaction is present for distances greater or equal to four hole diameters. Here, a spanwise distance of four hole diameters is chosen. Interestingly, the simulation of multiple holes leads to a notable degradation of the bleed performance. While the single-hole approach overpredicts the sonic flow coefficient, the single-row approach better fits the experimental data. Thus, an even more pronounced difference between the single- and multi-hole approach is assumed for larger porosities.

Moreover, two parameter effects are apparent in experimental and numerical data. First, the thickness-to-diameter ratio significantly affects the sonic flow coefficient. The lower the ratio, the higher the coefficient. For $T/D=2.0$, the data fits the regression of~\citet{Slater2012}, while the data for lower ratios lies below this regression. The second effect is the impact of the hole diameter. The upper two curves (plates 101 and 1501) represent holes with an equal thickness-to-diameter ratio. However, the sonic flow coefficient is higher for small hole diameters. Regarding the spanwise interaction of the holes, it is assumed that the effect would have been even more visible if the experiments were repeated with only one hole.

\subsection{Multi-hole validation}

\begin{figure}[b!]
    \centering
    \includegraphics[]{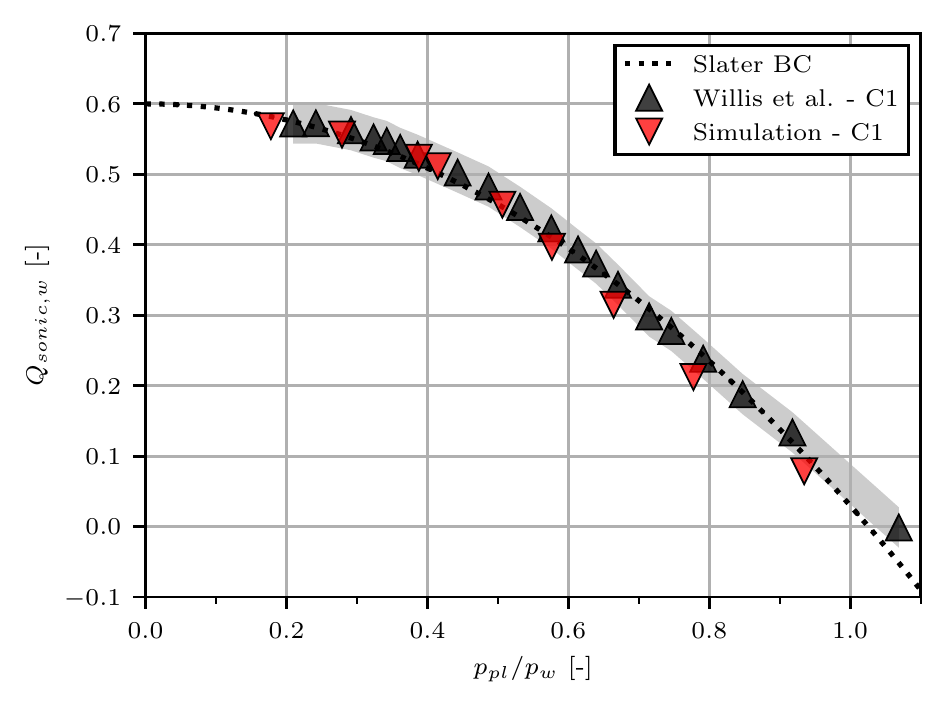}
    \caption{Experimentally and numerically determined surface sonic flow coefficient for the C1 plate of~\citet{Willis1995b} compared to the regression of~\citet{Slater2012}; gray area illustrates \SI{5}{\%} range around the experimental data}
    \label{fig:validationWillis}
\end{figure}

The C1 plate from the experiments of~\citet{Willis1995b} has a hole diameter $D=$ \SI{6.35}{\mm}, a thickness-to-diameter ratio $T/D=1$, a stagger angle $\beta=$ \SI{30}{\degree}, and a porosity $\phi=$ \SI{19.63}{\%}. The numerical setup consists of three columns and three rows of holes.

Again, the numerically obtained data fits very well with the experiments. The overall simulation trend is equal, as apparent in Fig.~\ref{fig:validationWillis}. For low pressure ratios $p_{pl}/p_w<0.6$, the simulations perfectly fit the data of~\citet{Willis1995b}. In contrast, the simulations slightly underpredict the sonic flow coefficients for larger ratios. However, the experimental data in this range should be regarded with caution since the point of intersection with the abscissa is at $p_{pl}/p_w \approx 1.05$.

Altogether, the numerical simulations show the same trend for both single- and multi-hole simulations as in the experiment. Thus, the numerical setup can be used confidently for a thorough parametric study.


\section{Results of the parametric study} \label{sec:results}

\begin{figure}[b!]
    \centering
    \begin{subfigure}[t!]{0.99\textwidth}
        \centering
        \hspace{15mm}
        \includegraphics[trim={0.mm 3.2cm 0.mm 0.3cm},clip]{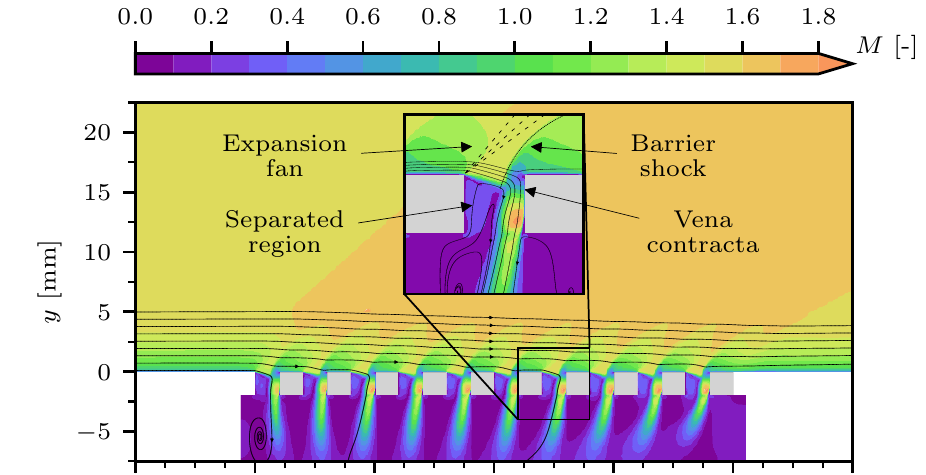}
        \caption{Hole-cutting plane}
        \label{fig:sliceSupersonicSym}
    \end{subfigure}
    \begin{subfigure}[t!]{0.99\textwidth}
        \centering
        \hspace{15mm}
        \includegraphics[trim={0.mm 0.3cm 0.mm 5.7cm},clip]{JAIAAtopologySlice.pdf}
        \caption{Offset-plane between holes}
        \label{fig:sliceSupersonicCen}
    \end{subfigure}

    \begin{tikzpicture}[remember picture,overlay]
        \node[xshift=-5.5cm, yshift=5.50cm] {\includegraphics[width=0.20\linewidth, trim={17mm 15mm 12mm 36mm},clip]{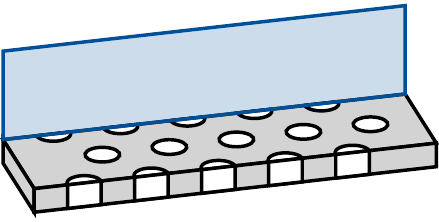}};
        \node[xshift=-5.5cm, yshift=2.15cm] {\includegraphics[width=0.20\linewidth, trim={17mm 15mm 12mm 36mm},clip]{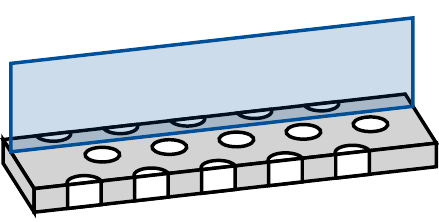}};
    \end{tikzpicture}
    \caption{Mach number field for a supersonic turbulent boundary bleeding; gray patches illustrate the out-of-plane holes}
    \label{fig:sliceSupersonic}
\end{figure}

CFD simulations have been performed for different plate lengths, hole diameters, porosity levels, thickness-to-diameter ratios, and stagger angles over a range of throat ratios to investigate the parameter influence on the mass flow rate and the supersonic turbulent boundary layer. An example of the flow field inside the bleed holes, the cavity plenum, and externally of the bleed is presented in Fig.~\ref{fig:sliceSupersonic}. The configuration has a plate length of $L_w=$ \SI{40}{\mm}, a hole diameter of $D = $ \SI{2}{\mm}, a thickness-to-diameter ratio of $T/D = 1$, a porosity of $\phi = $ \SI{22.67}{\percent}, and a stagger angle $\beta=$ \SI{60}{\degree}.

The flow field on the plane that cuts the holes is shown in Fig.~\ref{fig:sliceSupersonicSym}. The gray patches show the out-of-plane holes that are located at another streamwise position as a result of the hole staggering with the hole columns. A strong impact of the bleed holes on the flow inside the boundary layer is apparent, where the mass removal induces expansion fans, which, in turn, cause a bending of the flow towards the wall. The holes capture the flow close to the wall, which is sucked into the cavity.

The zoom-in in Fig.~\ref{fig:sliceSupersonicSym} illustrates the typical flow structure inside a bleed hole for a supersonic (upstream) bleed. An expansion fan is provoked at the leading edge of the hole, which redirects the flow into the hole. Downstream, the flow is decelerated as the hole wall at the rear creates a back pressure, which results in the appearance of a shock. The so-called barrier shock penetrates, like the expansion fan, into the boundary layer. As a result, the flow inside the boundary layer passes several expansions and compression waves. Downstream of the shock, the flow is redirected towards the hole exit, and the low pressure inside the plenum causes an acceleration of the flow to supersonic conditions. An under-expanded jet is clearly visible in the plenum. The mass flow rate passing the bleed hole is limited by the minimum size of the stream-tube, the so-called vena contracta area. Its position is roughly at the imaginary intersection of the barrier shock with the (lower) outer streamline of the captured flow. Flow separation is apparent near the front of the hole, where the flow enters the hole from the plenum side.

Fig.~\ref{fig:sliceSupersonicCen} shows the flow field on the plane between two columns of holes where the wall is fully solid. Still, the effect of the expansion fans and barrier shocks provoked by the holes is evident. The first hole generates a particularly distinctive expansion fan associated with a strong flow deflection. Downstream, the out-of-plane effects of the holes are apparent. The flow passes several expansion and compression waves induced by the flow into the holes located out of the plane. Moreover, a trailing shock at the end of the plate is prominent, which is the barrier shock of the last hole, redirecting the flow into the wall-parallel direction. The slope of the streamlines clearly demonstrates the thinning of the boundary layer.

The following subsections explain the influence of the different parameters on bleed efficiency and effectiveness. At any time, only one parameter is altered to isolate its effect. The default plate has a length $L_w=$ \SI{40}{\mm}, a hole diameter $D=$ \SI{0.5}{\mm}, a porosity $\phi=$ \SI{22.67}{\%}, a thickness-to-diameter ratio $T/D=1$, and a stagger angle $\beta=$ \SI{60}{\degree}, which is equal to the evenly described case but with a smaller hole diameter.


\subsection{Plate length} \label{sec:LW}

The influence of the plate length (or size) on efficiency and effectiveness has never been investigated to the author's knowledge. \citet{Syberg1973b} stated that the optimal size is given by the choking conditions. A choked plate removes the highest possible mass flow rate by simultaneously having the smallest possible plate size, which reduces roughness losses. However, no study compares two plates of different sizes with identical hole patterns.

\begin{figure}[t!]
    \centering
    \includegraphics[trim={0.cm 0.3cm 0.cm 0.3cm},clip]{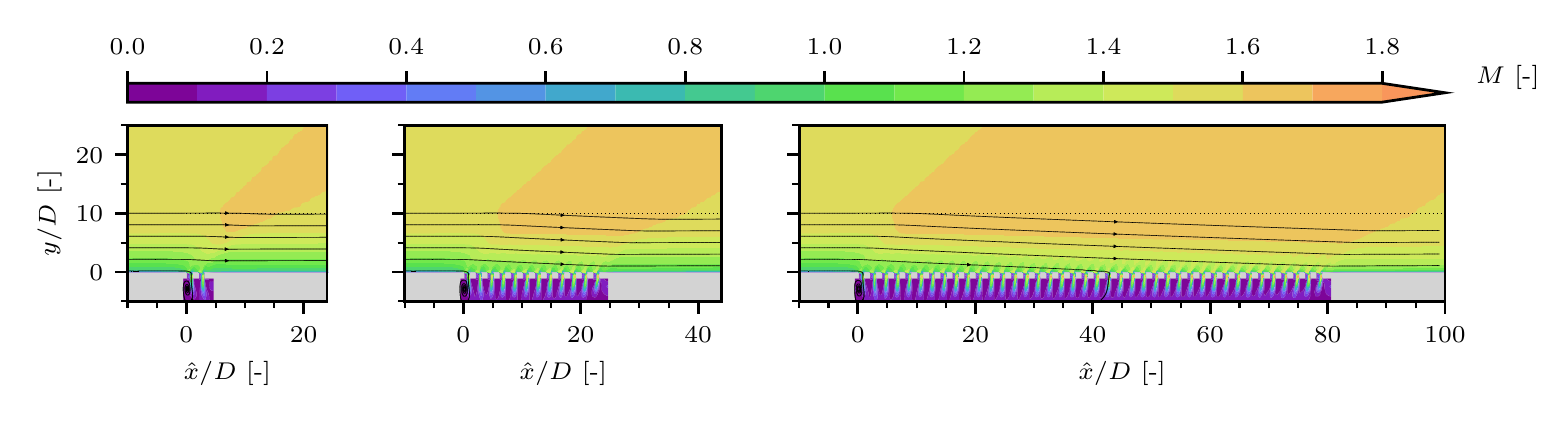}
    \begin{subfigure}[t!]{0.21\textwidth}
        \centering
        \caption{$L_w=$ \SI{2}{\mm}}
    \end{subfigure}
    \begin{subfigure}[t!]{0.23\textwidth}
        \centering
        \caption{$L_w=$ \SI{12}{\mm}}
    \end{subfigure}
    \begin{subfigure}[t!]{0.48\textwidth}
        \centering
        \caption{$L_w=$ \SI{40}{\mm}}
    \end{subfigure}
    \caption{Mach number contours for different plate lengths ($TR=0.7$); dotted line refers to the wall distance of a streamline outside the boundary layer at the beginning of the plate}
    \label{fig:ContoursLW}
\end{figure}

In this study, several plate lengths are examined. Fig.~\ref{fig:ContoursLW} shows the flow field along three plates with $L_w=$ \SIlist{2;12;40}{\mm}. The throat ratio $TR=0.7$ (see Eq.~\ref{eq:throatRatio}) is kept constant for the simulation to illustrate similar working regimes. Independently of the size, a thinning of the boundary layer is notable. However, the effect increases with longer plates as more mass is removed since the bleed area is larger than for short plates with fewer holes. The bending of the streamlines is illustrated by the dotted line, which serves as a reference indicating the upstream location of a streamline outside the boundary layer. Thus, the deflection is equal to those at the boundary layer edge.

\begin{figure}[htb!]
    \centering
    \includegraphics[trim={0.cm 0.3cm 0.cm 0.3cm},clip]{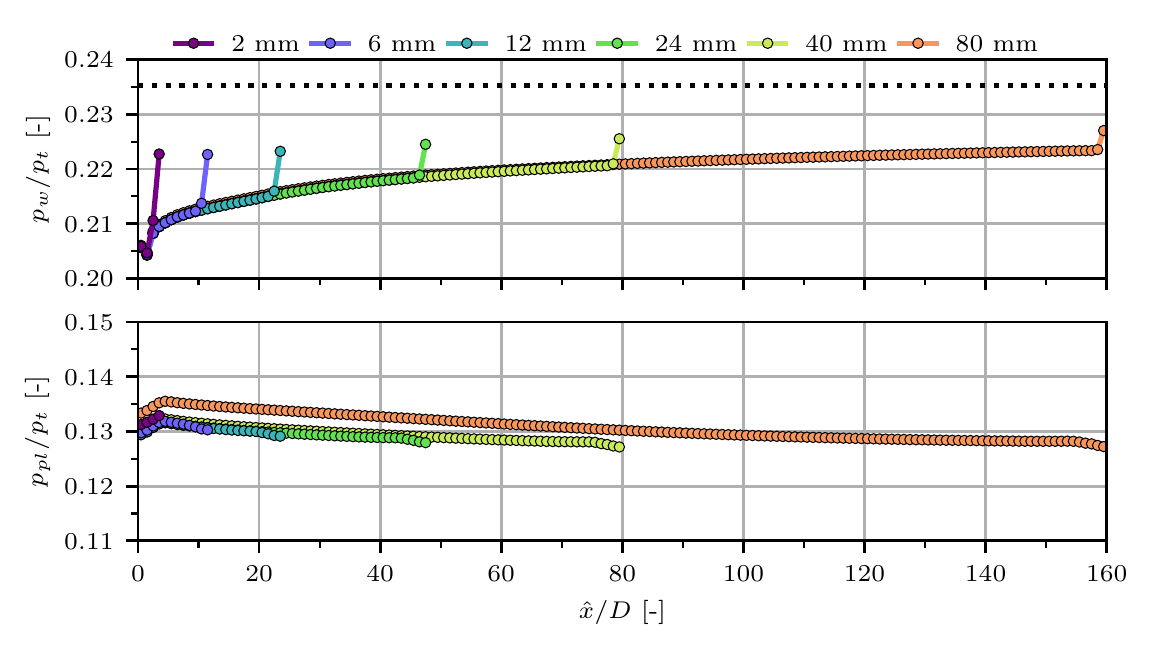}
    \caption{Static external wall and plenum pressure for different plate lengths ($TR=0.7$); the dotted line highlights the isentropic pressure ratio for $M=1.6$}
    \label{fig:PressureLW}
\end{figure}

The trend of the static wall and plenum pressure along the porous plates are detailed in Fig.~\ref{fig:PressureLW}. Independently of the plate length, the trend for the wall pressure is the same. At the beginning of the plate, the pressure drops because of the expansion fan. Along the plate, the pressure increases because of continuous compression caused by the barrier shocks and the steady deflection in wall-parallel directions. However, the slope decreases with further distance to the plate beginning. For the last hole, the wall pressure is higher because of the presence of the trailing shock. Moreover, the isentropic pressure ratio for $M=1.6$ is illustrated to pinpoint that the wall pressure is significantly lower than the calculated wall pressure, leading to higher ratios $p_{pl}/p_w$.

Furthermore, the trend for the plenum pressure is shown in the lower graph. Its value contrasts with the wall pressure since it is decreasing along the plate, caused by increasing pressure losses as the momentum of the sucked flow increases. Thus, the shock waves inside the holes become more intense. For longer plates, the plenum pressure is shifted to higher values at the beginning of the plate. In contrast, its average value stays approximately the same due to the choked throat. The higher the losses, the lower the plenum pressure, leading to a lower density and hence a lower mass flow rate for choking conditions. Consequently, the plenum pressure increases until an equilibrium between the sonic mass flow rate and the bleed mass flow rate is achieved.

\begin{figure}[htb!]
    \centering
    \includegraphics[trim={0.cm 0.2cm 0.cm 0.2cm},clip]{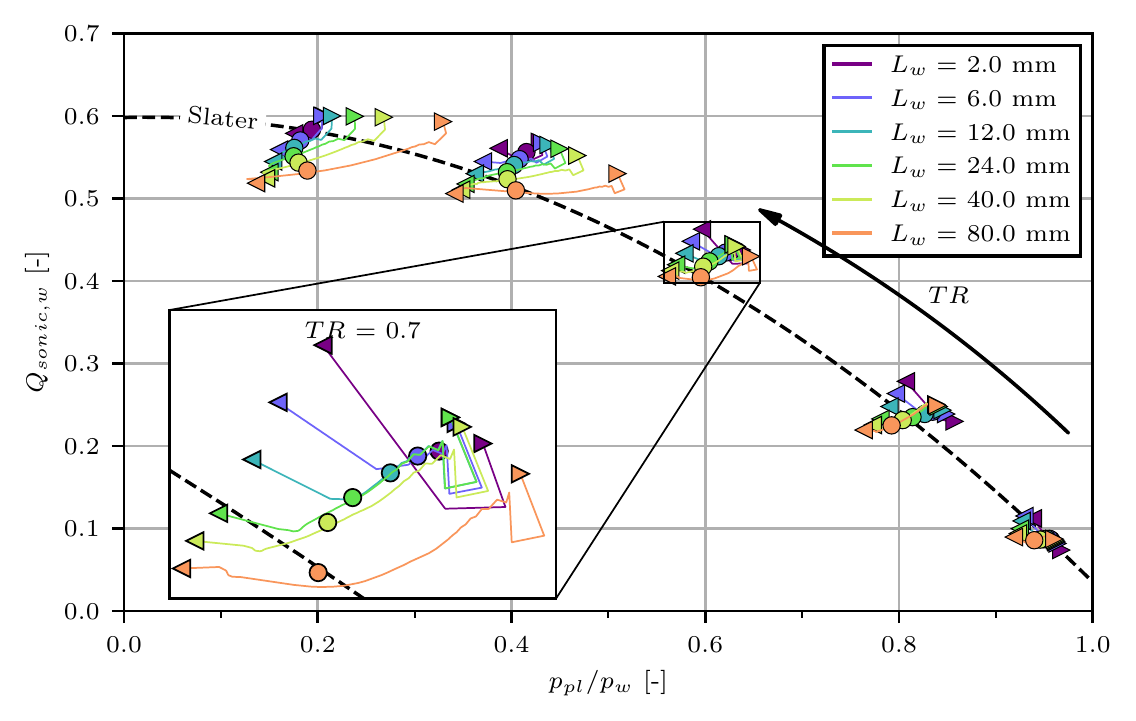}
    \caption{Surface sonic flow coefficient for different plate lengths}
    \label{fig:CaneLW}
\end{figure}

More simulations are performed for different throat ratios. Fig.~\ref{fig:CaneLW} shows the local surface sonic flow coefficient, which illustrates the bleed efficiency for different plate lengths as function of the pressure ratio from the plenum to the external wall. One solid line represents one simulation with one throat ratio. The right triangle ($\blacktriangleright$) illustrates the mass flow rate associated with the first hole, while the left triangle ($\blacktriangleleft$) stands for the last hole and the circle ($\bullet$) for the averaged value, i.e., the global plate working regime. Furthermore, the regression of~\citet{Slater2012}, which is not sensitive to any parameter variation, is shown as a reference.

A closer look at Fig.~\ref{fig:CaneLW} reveals essential differences between the plates for pressure ratios $p_{pl}/p_w<0.9$. For the first holes, the trend is similar. Along the plate, the pressure ratio is decreasing, which is a consequence of the increasing external wall pressure and higher total pressure losses induced by the shocks at the hole entry and inside the holes. Thus, the effect is the strongest for long plates and high throat ratios. Simultaneously, the sonic flow coefficient decreases due to the thinning of the boundary layer as the momentum of the sucked air increases, and thus, the size of the separated region inside the holes. Consequently, the momentum of the flow in the vicinity of the wall increases, and hence, the intensity of the barrier shock.

Since longer plates consist of more bleed holes, the negative effects induced by the barrier shocks become stronger. Thus, the global sonic flow coefficient decreases with the plate length. Another trend in the plot is the increase of the sonic flow coefficient at the last hole. The reason for that is the higher pressure around the last hole caused by the adverse pressure gradient of the trailing shock. The effect is weaker for long plates, as the boundary layer thinning is asymptotic, and the main flow direction bends successively back in the wall-parallel direction.

If the holes are choked, the deviation between the plates is more significant. For low pressure ratios, the sonic flow coefficient is around \SI{10}{\%} higher for a \SI{2}{\mm} plate compared to an \SI{80}{\mm} plate. However, the total bleed mass flow rate is way lower. On the contrary, no significant differences are apparent for high pressure ratios $p_{pl}/p_w>0.9$ as the holes are far from choking conditions.

\begin{figure}[b!]
    \centering
    \includegraphics[trim={0.cm 0.cm 0.cm 0.2cm},clip]{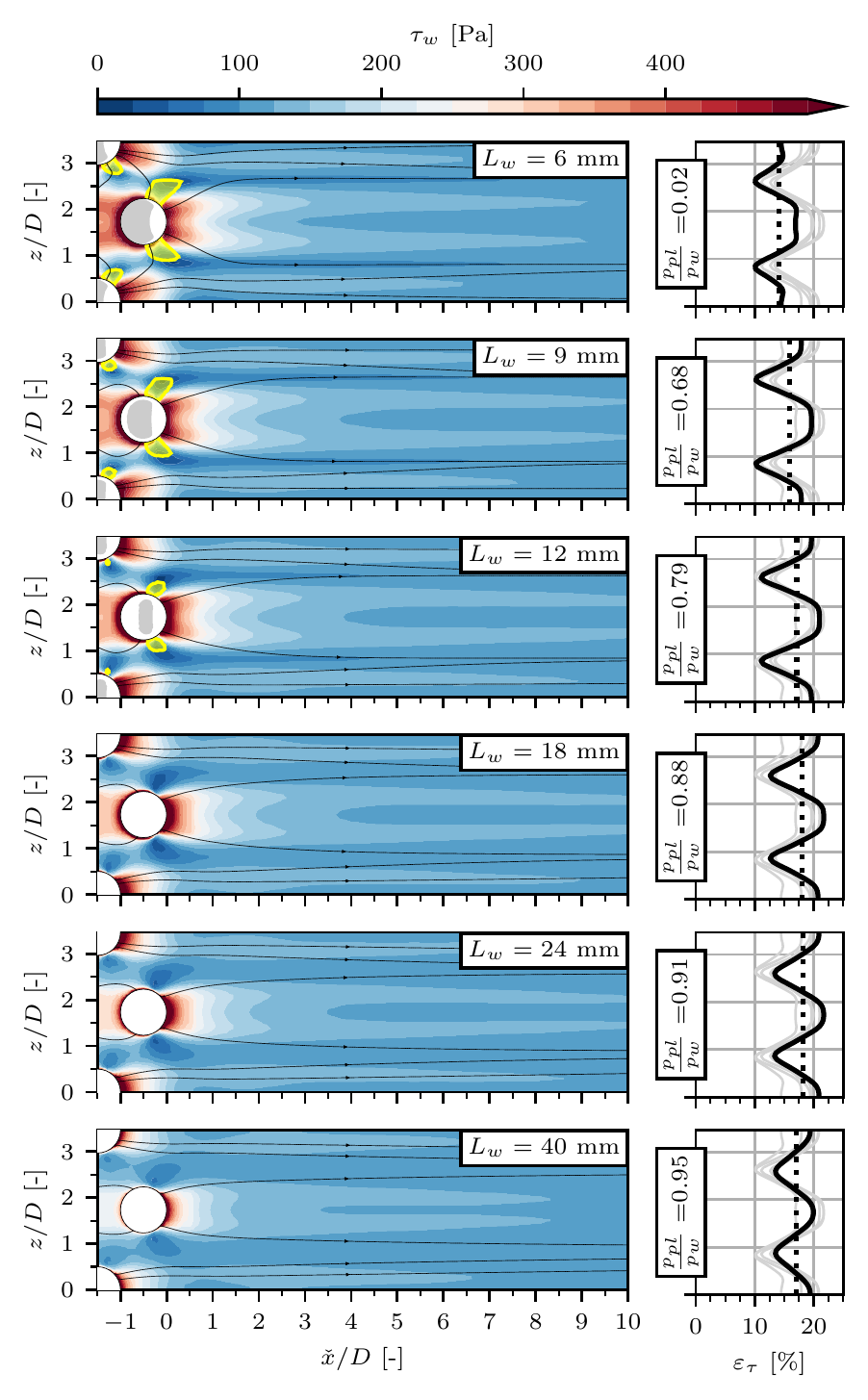}

    \begin{subfigure}[t!]{0.39\textwidth}
        \centering
        \caption{Nominal wall shear stress}
        \label{fig:shearLWnominal}
    \end{subfigure}
    \begin{subfigure}[t!]{0.25\textwidth}
        \centering
        \caption{Rise in wall shear stress}
        \label{fig:shearLWincrease}
    \end{subfigure}
    \caption{Comparison of the wall shear stress downstream of the bleed region for different plate lengths; yellow areas in (a) visualize regions with negative streamwise component and gray areas indicate supersonic hole entry flow; (b) wall shear stress \SI{10}{\mm} ($20 D$) downstream of the last hole compared to its value \SI{10}{\mm} upstream of the first hole; dotted line illustrates averaged value along the spanwise direction}
    \label{fig:shearLW}
\end{figure}

In the second step, simulations with constant bleed mass flow rates are performed to evaluate the effectiveness of the plates. The removed mass flow rate corresponds to \SI{25}{\%} of the displacement mass flow rate of the incoming boundary layer. In Fig.~\ref{fig:shearLW}, the wall shear stresses downstream of the bleed region are compared for plate lengths from \SIrange{6}{40}{\mm} for an equal bleed mass flow rate which varies between minimum and maximum by less than \SI{2}{\%}. From top to bottom, the plate length and the pressure ratio $p_{pl}/p_w$ increase. While the holes for the smallest plate are choked ($p_{pl}/p_w \le 0.528$), choking conditions are not achieved for the other plates.

Significant differences in the flow field around and downstream of the last holes are apparent in Fig.~\ref{fig:shearLWnominal}. The shorter the plates, the more significant the rise in the wall shear stress at the rear edge of the holes, caused by the higher suction rate through the hole. Moreover, in the case of short plates, the flow at the hole entry is partially supersonic (gray patches), resulting in a barrier shock inside the holes. Besides, the high suction rates lead to local flow reversal, visualized by the yellow patches. These areas grow with the suction rate and are only apparent for plates shorter than $L_w=$ \SI{18}{\mm}. As a result of the reversed flow, vortices are generated that cause an area of low wall shear stress further downstream. The friction lines highlight how the air streams in low-momentum areas where the flow is directed upwards. The smaller the plate and the pressure ratio, the more significant this effect. For the longest plate, the friction lines are significantly less bent.

Fig.~\ref{fig:shearLWincrease} presents the relative rise in the wall shear stress \SI{10}{\mm} downstream of the plate compared to the value \SI{10}{\mm} upstream of the first hole. Independently of the pressure ratio, a spanwise variation of the wall shear stress is notable, with its highest values located in the slipstream of the holes and the lowest values at the spanwise position where no holes are located. The dotted line highlights the mean value, while the gray lines present the curve for the other lengths.

Again, the effect of the reversed flow areas is apparent. For the three smallest plates, the rise in the wall shear stress between the holes is significantly lower and has a local value of approximately \SI{10}{\%} compared to the \SI{15}{\%} for the long plates. Hence, high suction rates negatively influence the effect of the porous bleed.

A second negative effect is visible for the three shortest plates, where the pressure downstream of the holes drops. The reason is the stronger adverse pressure gradient induced by the barrier shock caused by the partially supersonic flow in the hole entry. The more extended the supersonic area, the higher the Mach number upstream of the shock and, thus, the shock intensity. Moreover, the barrier shock moves closer to the rear edge, and some outflow is present~\cite{Oorebeek2015}. Especially for the shortest plate with a \SI{6}{\mm} length, the increase in the wall shear stress is significantly lower.

However, a too-long plate also results in a lower rise in the wall shear stress. A lower increase downstream of the holes is notable for the longest plate with $L_w=$ \SI{40}{\mm}. Apparently, the roughness effect play for low suction rates a more significant role. Altogether, the most intense increase of the wall shear stress is found for plates around \SI{20}{\mm}, corresponding to a pressure ratio of $p_{pl}/p_w \approx 0.9$.

\begin{figure}[htb!]
    \centering
    \includegraphics[trim={0.cm 0.cm 0.cm 0.cm},clip]{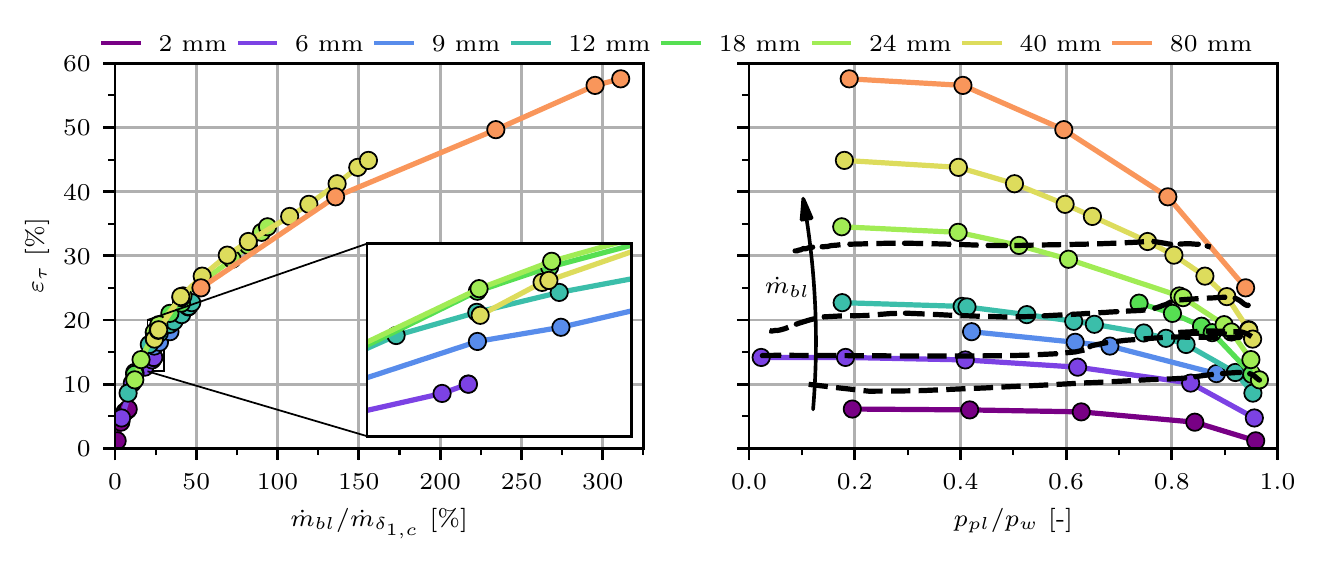}
    \begin{subfigure}[t!]{0.45\textwidth}
        \centering
        \caption{Rise in wall shear stress as function of the bleed mass flow rate}
        \label{fig:effectivenessLWm}
    \end{subfigure}
    \hspace{1mm}
    \begin{subfigure}[t!]{0.42\textwidth}
        \centering
        \caption{Rise in wall shear stress as function of the pressure ratio}
        \label{fig:effectivenessLWp}
    \end{subfigure}
    \caption{Rise in the wall shear stress as function of the bleed mass flow rate $\dot{m}_{bl}$ and the pressure ratio $p_{pl}/p_w$ for varying plate lengths; dashed lines in (b) pinpoint lines of constant bleed mass flow rate}
    \label{fig:effectivenessLW}
\end{figure}

All simulations performed for the standard plate for various bleed mass flow rates are evaluated regarding the effectiveness and shown in Fig.~\ref{fig:effectivenessLW}. First, the rise in the wall shear stress is illustrated over the relative bleed mass flow rate (see Fig.~\ref{fig:effectivenessLWm}). The higher the sucked mass, the more effective the porous bleed. However, large plates are required for large mass flow rates since choking limits the maximum removable mass flow rate. Moreover, the slope of the general trend decreases with the plate length and bleed mass flow rate.

Interestingly, by observing the effect as function of the pressure ratio, as shown in Fig.~\ref{fig:effectivenessLWp}, an optimum can be located for each bleed mass flow rate (dashed lines) around $p_{pl}/p_w \approx 0.9$. Contrary to~\citet{Syberg1973b}, choking conditions decrease the effect of the porous bleed and are more critical than the roughness induced by the holes. With the choking of the holes, the wall shear stress downstream of the bleed region decreases again. Nevertheless, it must be stated that too high pressure ratios may lead to undesired behavior (e.g., blowing) if the inflow conditions change and the external wall pressure drops below the plenum pressure.


\subsection{Hole diameter} \label{sec:D}

The second investigated parameter is the hole diameter. Multiple studies~\cite{Davis1997,Akatsuka2006,Eichorn2013} have shown a hole diameter influence. However, these effects are not considered in existing bleed models. In this study, the hole diameter is varied by more than one order of magnitude from $D=$ \SIrange{0.25}{4.00}{\mm}.

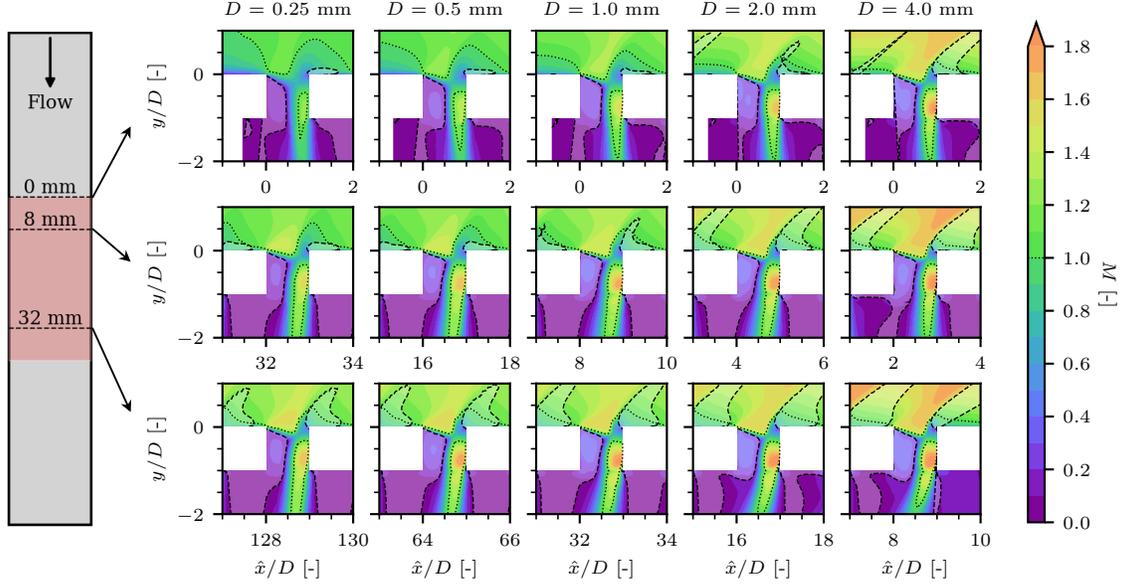
\begin{figure}[t!]
    \centering
    \begin{subfigure}[t!]{0.10\textwidth}
        \includegraphics[trim={0.cm 0.cm 0.cm 0.3cm},clip]{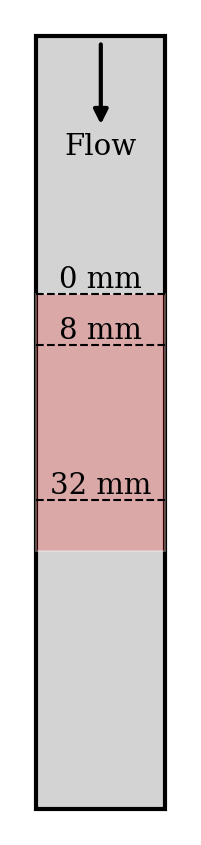}
    \end{subfigure}
    \begin{subfigure}[t!]{0.83\textwidth}
        \centering
        \includegraphics[trim={0.0cm 0.2cm 0.cm 0.2cm},clip]{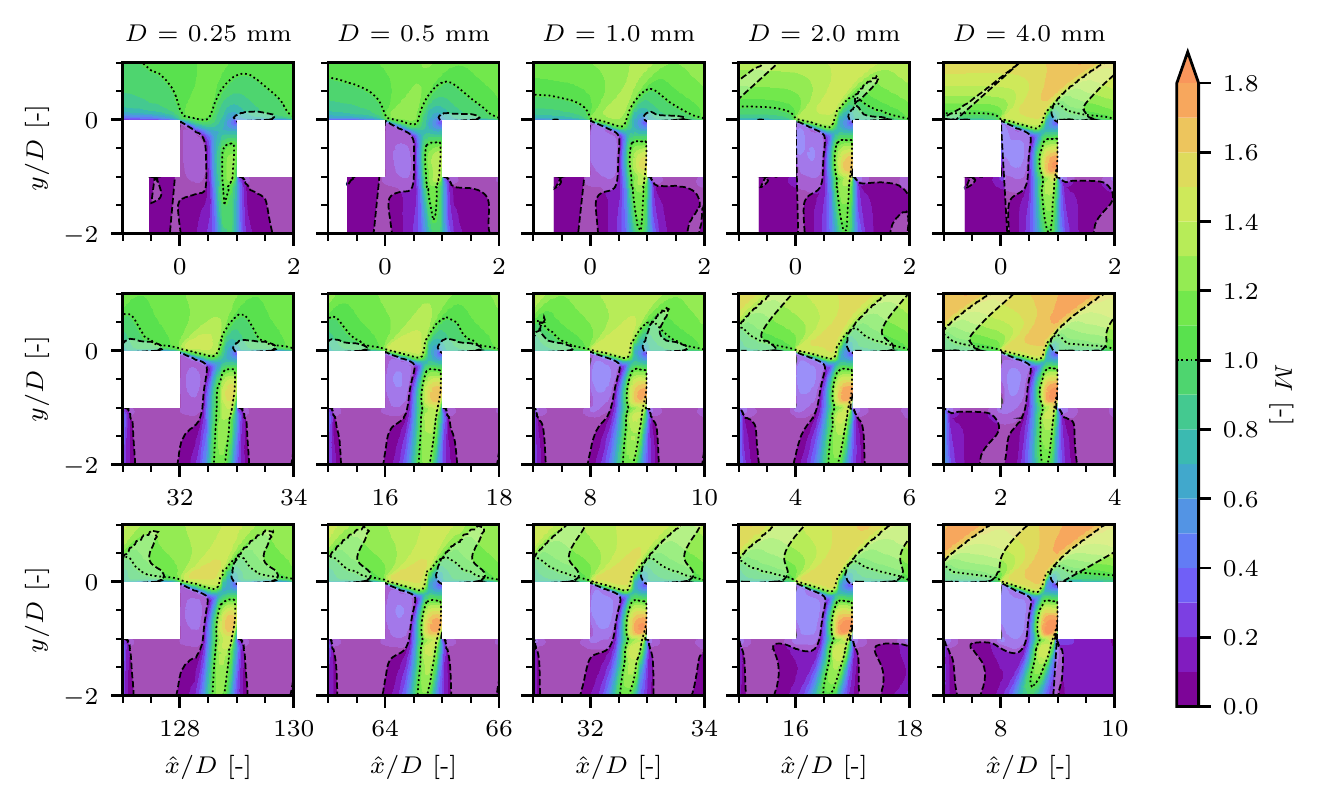}
    \end{subfigure}
    \caption{Mach number contours of the flow in the bleed holes for different hole diameters at $\hat{x}=$ \SIlist{0;8;32}{\mm} (top to bottom) for $TR=0.7$; dotted line illustrates sonic line; brightened areas outlined by dashed lines illustrate regions with positive wall-normal flow momentum}
    \label{fig:sliceZoomD}
    \begin{tikzpicture}[remember picture,overlay]
        \draw [black,-{Stealth[length=1.5mm, width=1.5mm]},line width=0.25mm] (-6.31,5.15) -- (-5.80,6.10);
        \draw [black,-{Stealth[length=1.5mm, width=1.5mm]},line width=0.25mm] (-6.31,4.74) -- (-5.80,4.30);
        \draw [black,-{Stealth[length=1.5mm, width=1.5mm]},line width=0.25mm] (-6.31,3.42) -- (-5.80,2.30);
    \end{tikzpicture}
\end{figure}

In the first step, the flow fields inside the holes are compared without altering the boundary layer thickness at different positions along the porous plate for a fixed throat ratio $TR=0.7$ and a length of \SI{40}{\mm}, as shown in Fig.~\ref{fig:sliceZoomD}. Both streamwise and wall-normal coordinates are normalized by the hole diameter to check similarity in the flow topology. From left to right, the hole diameter increases, and going from top to bottom corresponds to a shift in the streamwise direction. The different positions are illustrated in the schematic on the left and correspond to \SIlist{0;20;80}{\%} of the plate length.

The sonic line is added as a dotted line to the plots as a characteristic indicator of the boundary layer fullness. Observing the sonic line in the first row reveals significant differences in the flow field for different hole diameters. The smaller the hole diameter, the lower the captured momentum since the maximum mass flow rate passing the hole is fixed by the hole area. Thus, small holes enable the sole removal of flow in the near-wall region with low flow momentum. For the maximum diameter of $D = $ \SI{4}{\mm}, the sonic line is relatively close to the wall, illustrating the high momentum in the vicinity of the wall. A strong expansion fan is observed, and the maximum Mach number inside the hole is higher than for smaller diameters. As a consequence, the barrier shock increases in intensity and moves downstream. Thus, higher pressure losses are expected and consequently a lower bleed efficiency.

As a second tool to compare the flow fields, the areas with a positive wall-normal momentum are brightened and outlined by dashed lines. Two significant differences are apparent between the different diameters in the first row. First, a flow in the upward direction downstream of the rear edge of the holes is visible. This is caused by the stagnation point inside the hole, causing a local blowing into the boundary layer in the rear part of the hole. As a result, the boundary layer thickens locally, which is an undesired effect. The higher the momentum of the flow and the shock intensity of the barrier shock, the larger the upstream flow area. Thus, the larger holes have a negative effect on the boundary layer thinning.

Moreover, the upstream flow inside the holes indicates the size of the present flow separation, which limits the vena contracta area, and hence, the bleed efficiency. For all hole diameters, the separated region inside the first hole is of similar size. However, the vena contracta area is slightly more constricted for large diameters. Furthermore, the reversed flow has a higher Mach number indicating a higher momentum and more losses.

The second row shows the holes at \SI{20}{\%} of the porous plate length. The flow fields are more similar in all cases, as seen by the position of the sonic line upstream of the hole. The main difference is the shock intensity since the upstream Mach number is again a function of the hole size. Also, the Mach number in the core of the supersonic jet streaming into the plenum is enlarged for increasing diameters. Thus, the losses inside the cavity caused by the under-expanded jet are higher and decrease the efficiency. Moreover, the higher shock intensity leads to a higher wall-normal flow momentum downstream of the hole, independently of the hole size. Thus, all negative effects are enhanced, which goes in line with the findings from Sec.~\ref{sec:LW}.

In the last row, the flow fields look self-similar for all diameters. The trend of the sonic line is equal, and only the external Mach number is higher for large diameters, which is a result of the higher momentum of the removed flow. However, the areas with a positive wall-normal flow momentum are significantly smaller for small holes, and also the momentum of the backflow inside the holes.

Comparing the changes of the flow field along the plate, a stronger thinning effect is observed for small holes. While the distance of the sonic line to the wall remains constant for large diameters, it moves closer to the wall for small hole diameter. As a consequence, a higher effectiveness is expected for small holes.

\begin{figure}[htb!]
    \centering
    \includegraphics[trim={0.cm 0.2cm 0.cm 0.2cm},clip]{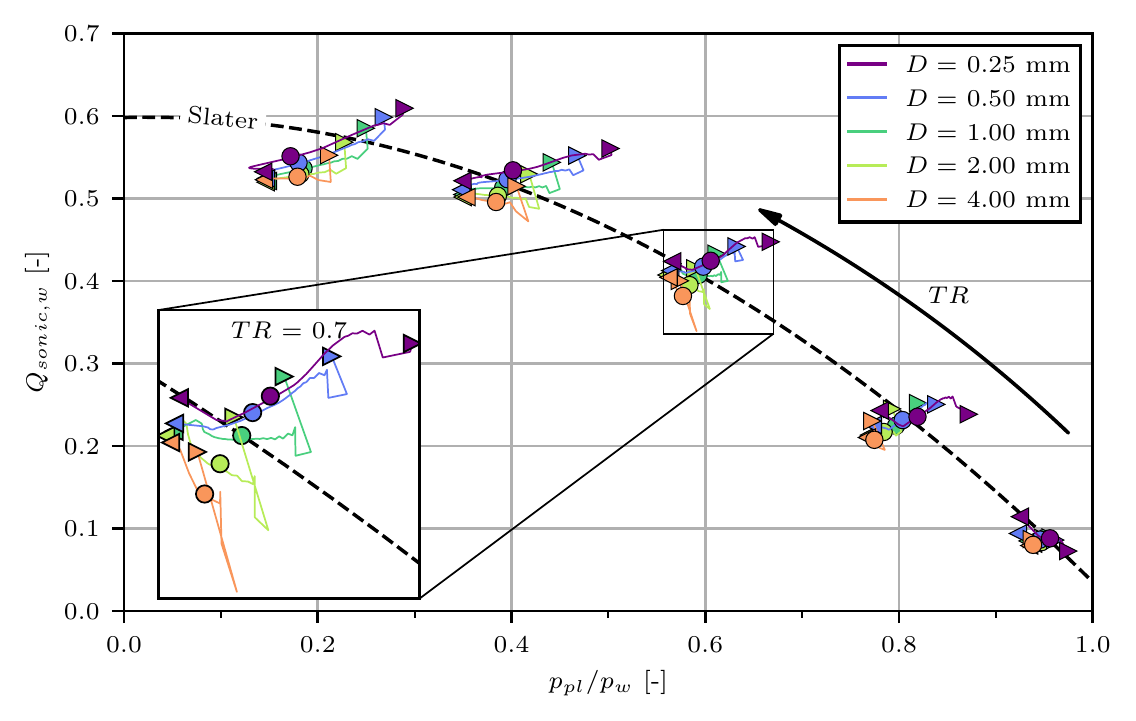}
    \caption{Surface sonic flow coefficient for different hole diameters}
    \label{fig:CaneD}
\end{figure}

The influence of the hole diameter on the bleed efficiency is illustrated in Fig.~\ref{fig:CaneD}, where the sonic flow coefficient as function of the pressure ratio $p_{pl}/p_w$ is shown. As stated in the last section, the wall pressure increases along the plate, which leads to a variation in the pressure ratio and a change in the sonic flow coefficient. The smaller the holes, the more important the variation of the pressure ratio since the number of holes increases along with the thinning effect, leading to a more marked expansion fan at the beginning of the plate.

Observing the first holes ($\blacktriangleright$), significant differences are apparent. As shown in Fig.~\ref{fig:sliceZoomD}, the flow into the holes differs as the sucked flow has a lower momentum. The vena contracta area is more extensive, which increases the removed mass flow rate and, as a result, the sonic flow coefficient. Moreover, the pressure ratio is higher as the main flow deflection caused by the boundary layer thinning is more distinct. The higher the throat ratio, the more pronounced the deviation between the different curves.

Interestingly, the differences for the last holes ($\blacktriangleleft$) are minor, which can be linked to the self-similarity found in Fig.~\ref{fig:sliceZoomD}. As the sucked flow momentum increases along the plate, the sonic flow coefficient decreases as the vena contracta area become smaller. Moreover, the positive slope of the pressure along the plate is higher for small hole diameters resulting in similar conditions for the last holes. However, significant differences in the global flow coefficient ($\bullet$) for the plates are visible: the smaller the holes, the higher the sonic flow coefficient and the pressure ratio. This effect increases with choking conditions. Our results are in line with the previous results of~\citet{Eichorn2013} and confirm the influence of the hole diameter on the sonic flow coefficient even for plates with more than one hole.

\begin{figure}[htb!]
    \centering
    \includegraphics[trim={0.cm 3.4cm 0.cm 2.8cm},clip]{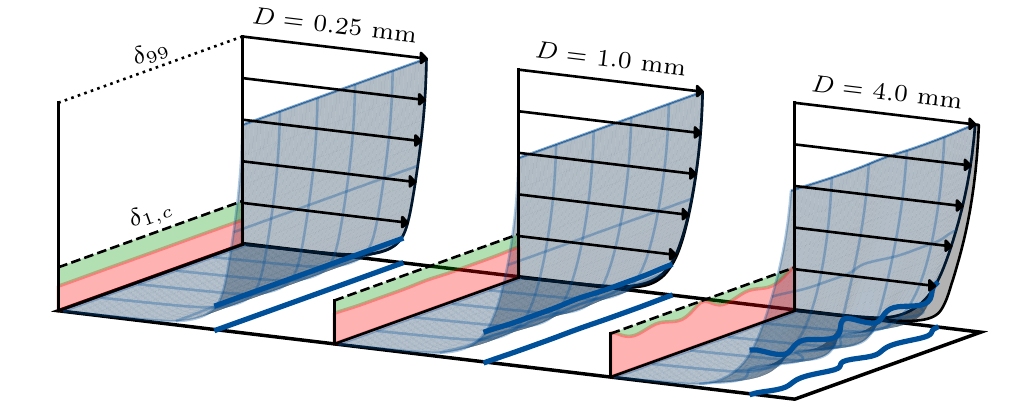}
    \caption{Boundary layer \SI{10}{\mm} downstream of the bleed region for $D=$ \SIlist{0.25;1.00;4.00}{\mm} (left to right); gray patch illustrates the envelope of the boundary layer profiles along the spanwise direction; red area details the compressible displacement thickness and green area the difference to the inflow}
    \label{fig:profiles3DD}
\end{figure}

Next, the boundary layer profiles downstream of the bleed region are discussed in order to stress on the effectiveness. \citet{Davis1997} measured significantly fuller boundary layer profiles using microporous plates compared to the C1 plate used by~\citet{Willis1996}. Thus, an increase in the effectiveness is expected for small hole diameters. Fig.~\ref{fig:profiles3DD} shows the boundary layer profiles along the span \SI{10}{\mm} downstream of the last hole for three different hole diameters. Independently of the hole diameter, the bleed mass flow rate is set so that \SI{107}{\%} of the incoming displacement mass flow rate is removed. All profiles are normalized by the free-stream streamwise velocity. Moreover, the boundary layer thickness $\delta_{99}$ extracted \SI{10}{\mm} upstream of the holes is used to normalize the wall-normal direction. The boundary layer thickness is not locally extracted, as the trailing shock induced by the porous bleed may lead to an overestimation.

The gray patch in the back corresponds to the envelope of the boundary layer profiles along the spanwise direction. The larger the diameter of the holes, the less homogeneous the flow field along the span resulting in locally fuller profiles. The variation of the boundary layer shape factor is given in Tab.~\ref{tab:Hdiameter}. For all diameters, the shape factor is lower compared to the inflow. However, the profiles downstream of the plate with $D=$ \SI{0.25}{\mm} are significantly fuller, while using larger holes leads to local spots where no fuller profile is obtained.

\begin{table}[t!]
    \centering
    \begin{tabularx}{0.76\textwidth}{x{0.09\textwidth}x{0.07\textwidth}x{0.07\textwidth}x{0.09\textwidth}x{0.09\textwidth}x{0.09\textwidth}x{0.09\textwidth}}
         \toprule
         $D$ &$H_{min}$ &$H_{max}$ & $\delta_{1,c_{min}}$ & $\delta_{1,c_{max}}$ & $\varepsilon_{\tau_{min}}$ & $\varepsilon_{\tau_{max}}$ \\
         \midrule
         \SI{0.25}{\mm} &\multicolumn{2}{c}{1.25} &\multicolumn{2}{c}{\SI{0.49}{\mm}} & \SI{37.4}{\%} & \SI{43.7}{\%} \\
         \SI{1.00}{\mm} &1.27 &1.30 & \SI{0.59}{\mm} & \SI{0.61}{\mm} & \SI{18.7}{\%} & \SI{34.4}{\%} \\
         \SI{4.00}{\mm} &1.25 &1.35 & \SI{0.64}{\mm} & \SI{0.88}{\mm} & \SI{-9.7}{\%} & \SI{37.1}{\%} \\
         \midrule
         \textit{Inflow}  &\multicolumn{2}{c}{\textit{1.36}} &\multicolumn{2}{c}{\textit{\SI[detect-all=true]{0.87}{\mm}}} &\multicolumn{2}{c}{\textit{--}} \\
         \bottomrule
    \end{tabularx}
    \caption{Boundary layer shape factor, displacement thickness, and rise in the wall shear stress for the boundary layer profiles shown in Fig.~\ref{fig:profiles3DD}}
    \label{tab:Hdiameter}
\end{table}

Additionally, the variation of the (compressible) displacement thickness is shown in Fig.~\ref{fig:profiles3DD}. The dashed line indicates the inflow displacement thickness and the red area is the one extracted downstream of the bleed region. The area between is colored in green and pinpoints the improvement compared to the inflow. Again, the minimum and maximum are listed in Tab.~\ref{tab:Hdiameter}. We see a similar trend as for the shape factor with larger values and variations for big holes. For the $D=$ \SI{4}{\mm} holes, a local thickening of the boundary layer is apparent at some spanwise locations.

\begin{figure}[b!]
    \centering
    \includegraphics[trim={0.cm 7.0cm 0.cm 0.2cm},clip]{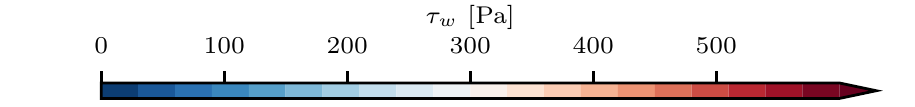}
    \includegraphics[trim={0.cm 0.2cm 0.cm 1.4cm},clip]{JAIAAshearContourD.pdf}

    \begin{subfigure}[t!]{0.39\textwidth}
        \centering
        \caption{Nominal wall shear stress}
        \label{fig:shearDnominal}
    \end{subfigure}
    \begin{subfigure}[t!]{0.25\textwidth}
        \centering
        \caption{Rise in wall shear stress}
        \label{fig:shearDincrease}
    \end{subfigure}
    \caption{Comparison of the wall shear stress downstream of the bleed region for different hole diameters; yellow areas in (a) visualize regions with negative streamwise component and gray areas indicate supersonic hole entry flow; (b) wall shear stress \SI{10}{\mm} downstream of the last hole compared to its value \SI{10}{\mm} upstream of the first hole; dotted line illustrates averaged value along the spanwise direction}
    \label{fig:shearD}
\end{figure}

The difference in the wall shear stress between the different hole diameters is visualized in Fig.~\ref{fig:shearD}. Again, the bleed mass flow rate is kept constant and corresponds to \SI{107}{\%} of the inflow displacement mass flow rate. Similar trends as for the boundary layer profiles are found: the smaller the hole, the higher the increase in the wall shear stress. Even though some local reversed flow is present close to the holes, the tiny holes lead to a significantly more homogeneous wall shear stress distribution downstream of the plate. This is a result of the lower shock intensity of the barrier shock caused by a lower Mach number at the hole entry. The rise in the wall shear stress is, on average, two times higher for $D=$ \SI{0.25}{\mm} compared to $D=$ \SI{4.00}{\mm}.

Moreover, the wall shear stress varies substantially along the spanwise direction for the largest holes. Downstream of the hole center on the hole-cutting plane, the wall shear stress is negatively affected at some spanwise positions as the higher intensity of the barrier shock leads to a local boundary layer thickening. Also, between the holes, the increase of the wall shear stress is significantly lower. On the other hand, local peaks are visible, which is in line with the trend for the boundary layer displacement thickness. Altogether, the rise in the wall shear stress is of the same order of magnitude as the decrease in the displacement thickness (see Tab.~\ref{tab:Hdiameter}), which underpins the use of this value to quantify the effectiveness. 

\begin{figure}[htb!]
    \centering
    \includegraphics[trim={0.cm 0.cm 0.cm 0.cm},clip]{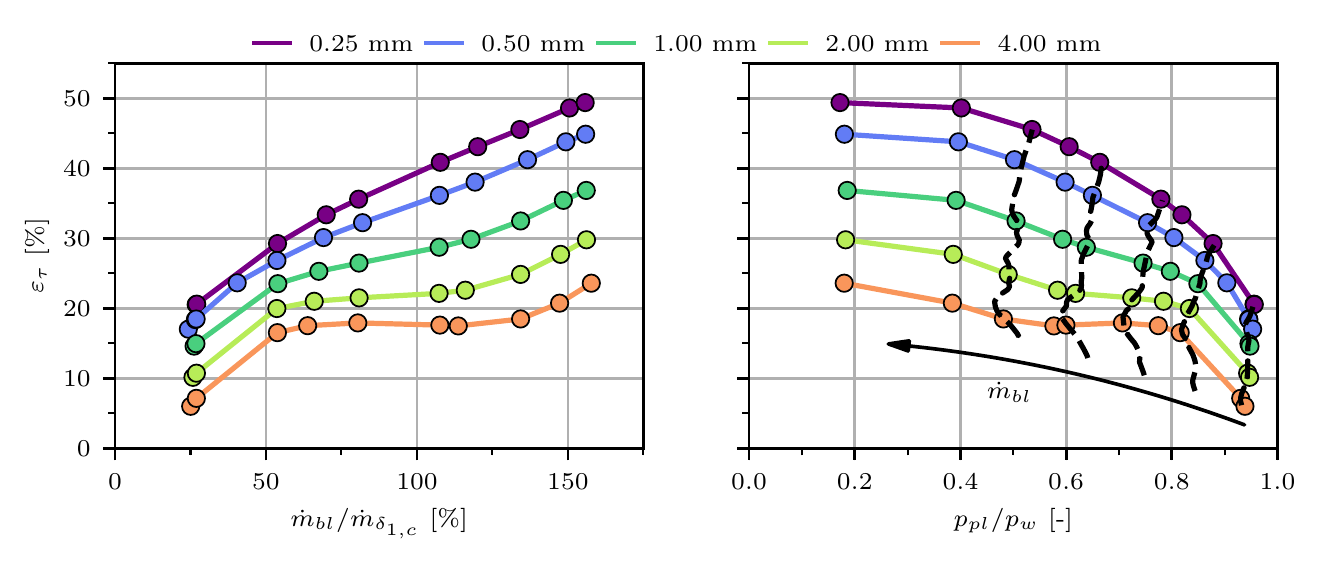}
    \begin{subfigure}[t!]{0.45\textwidth}
        \centering
        \caption{Rise in wall shear stress as function of the bleed mass flow rate}
        \label{fig:effectivenessDm}
    \end{subfigure}
    \begin{subfigure}[t!]{0.42\textwidth}
        \centering
        \caption{Rise in wall shear stress as function of the pressure ratio}
        \label{fig:effectivenessDp}
    \end{subfigure}
    \caption{Rise in the wall shear stress as function of the bleed mass flow rate $\dot{m}_{bl}$ and the pressure ratio $p_{pl}/p_w$ for varying hole diameters; dashed lines in (b) pinpoint lines of constant bleed mass flow rate}
    \label{fig:effectivenessD}
\end{figure}

In the last step, the rise in the wall shear stress is shown as function of the bleed mass flow rate and the plenum pressure in Fig.~\ref{fig:effectivenessD}. Similar to Fig.~\ref{fig:effectivenessLW}, higher wall shear stresses are found with an increase in air removal. However, the slope strongly depends on the hole diameter. The smaller the holes, the higher the slope. Thus, the effectiveness is notably affected by the hole diameter.

Moreover, by looking at Fig.~\ref{fig:effectivenessDp}, this effect is independent of the pressure ratio. By reducing the hole diameter, the wall shear stress can be significantly increased with keeping the bleed mass flow rate constant. Hence, using small holes is favorable with regard to both effectiveness and efficiency.


\subsection{Porosity} \label{sec:phi}

\begin{figure}[ht!]
    \centering
    \includegraphics[trim={0.cm 0.2cm 0.cm 0.2cm},clip]{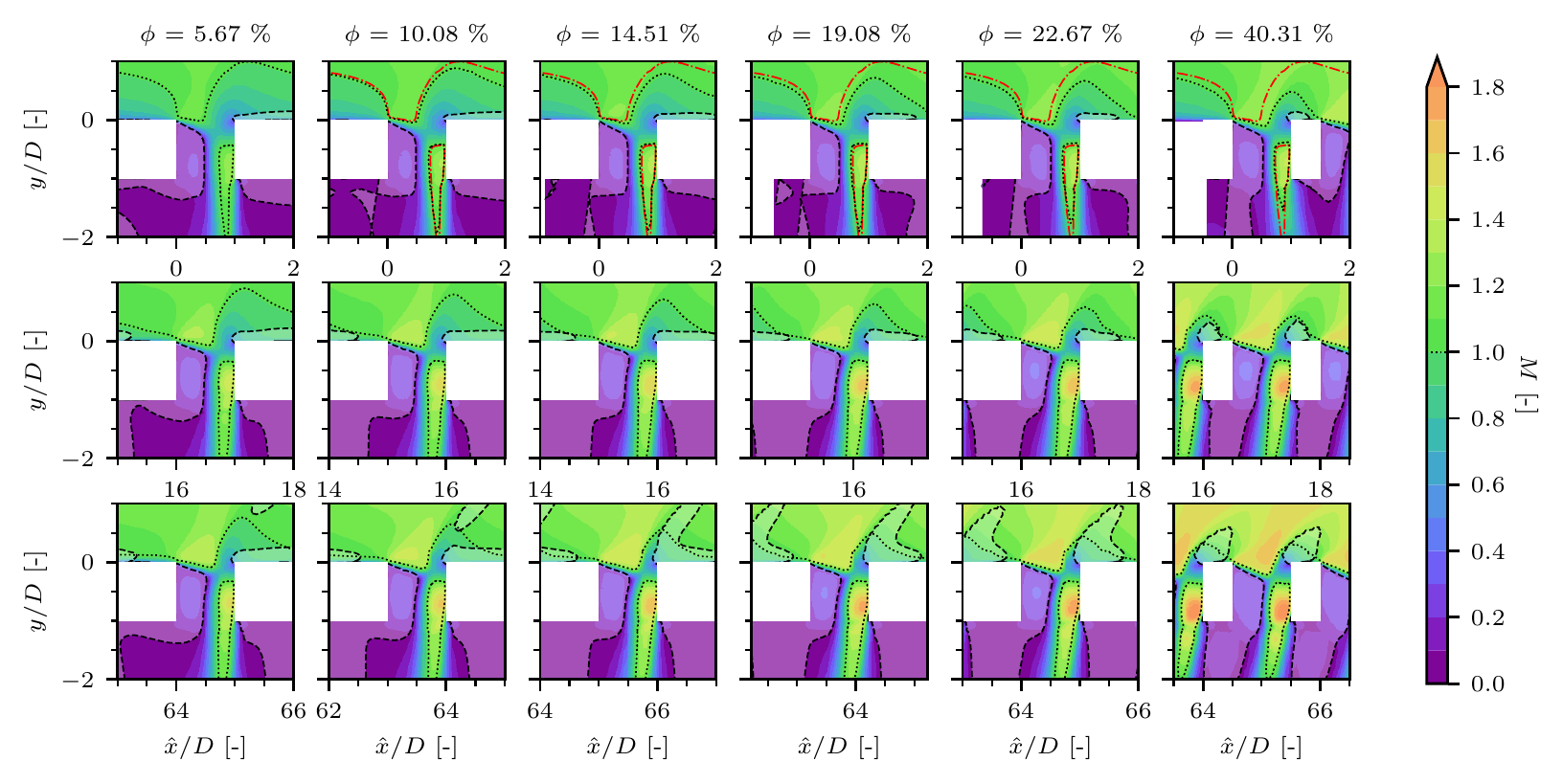}
    \caption{Mach number contours of the flow in the bleed holes for different porosity levels at $\hat{x}\approx$ \SIlist{0;8;32}{\mm} (top to bottom) for $TR=0.7$; dotted line illustrates sonic line; brightened areas outlined by dashed lines illustrate regions with positive wall-normal flow momentum}
    \label{fig:zoomPhi}
\end{figure}

The porosity, which describes the area fraction of holes in the solid wall, is of major importance in removing a required mass flow rate. Low porosity levels result in a lower removable amount of air as the total suction area is small with respect to the plate area. A porosity of \SI{100}{\%} corresponds to a bleed slot, which may have a lower length, but, consequently, also a more local impact on the flow. In this study, the porosity varies from $\phi=$ \SIrange{5.67}{40.31}{\%}. The flow field inside the holes is shown in Fig.~\ref{fig:zoomPhi} for all investigated porosity levels.

Independently of the porosity, the hole diameter remains constant ($D=$ \SI{0.5}{\mm}), which leads to the assumption of equal flow fields in the first holes as the inflow is supersonic, and hence, the upstream influence small. However, by looking at the first row of Fig.~\ref{fig:zoomPhi}, differences in the flow fields are apparent. The higher the porosity, the more prominent the interaction between the holes. This is visible by following the sonic line, which reveals the position of the barrier shock. As reference, the sonic line for the lowest porosity is illustrated by the red dash-dotted line. Even though the pressure ratio is similar, the position of the barrier shock is further downstream for high porosities. This is a result of the suction of the downstream located hole, which reduces the back pressure leading to a movement of the shock. A consequence of the further downstream lying barrier shock is a higher shock intensity and a higher Mach number of the flow inside the holes. Therefore, higher losses are expected.

A second effect of the hole interaction seen in the first row is the smaller longitudinal size of the region of an upward-directed momentum between the holes. The size of the area is limited by the expansion fan of the next hole, which is closer in case of high porosities. Thus, a more substantial thinning effect is expected. Also, the trend of the sonic line confirms this assumption as it is located significantly closer to the wall downstream of the holes for high porosities.

Further downstream on the plate, the same effects are present, as shown in the second row. The sonic line moves closer to the wall for high porosities. Moreover, the Mach number is higher, and thus the shock intensity is stronger.

In the last row, all effects are strengthened. The deflection of the flow inside the holes is the strongest for the highest porosity, which results in a significantly higher Mach number at the entry of the hole. The barrier shock is consequently stronger and located further downstream, with the losses expected to be more significant. Moreover, the sonic line is located closer to the wall, and the area of upward-directed momentum is smaller in size. Hence, higher effectiveness of large porosities is expected by having a lower efficiency simultaneously.

\begin{figure}[h!]
    \centering
    \includegraphics[trim={0.cm 0.2cm 0.cm 0.2cm},clip]{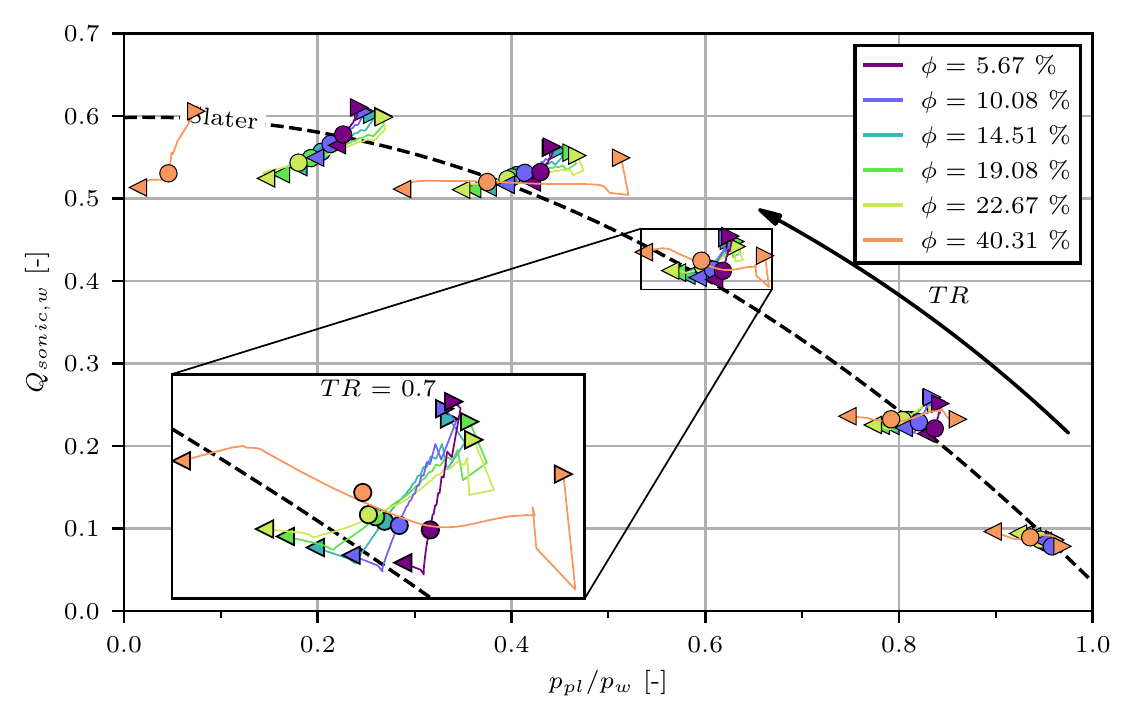}
    \caption{Surface sonic flow coefficient for different porosity levels}
    \label{fig:CanePhi}
\end{figure}

The locally extracted surface sonic flow coefficient is shown compared to the model of~\citet{Slater2012} in Fig.~\ref{fig:CanePhi}. A variation of the pressure ratio along the porous plate is apparent. The higher the porosity level, the higher the horizontal distance between the first and last holes.  This results from the higher pressure losses induced by stronger barrier shocks and the larger number of holes. Moreover, the effect is enhanced for large throat ratios where the drop in the total pressure along the x-position increases because of enhanced shock intensities with lower pressure ratios.

A second trend is the decrease of the sonic flow coefficient with the position along the plate. This behavior is a function of the porosity and increases for low porosity levels. The explanation can be found in Fig.~\ref{fig:zoomPhi}, where a constriction of the vena contracta area is apparent for the downstream lying holes. Interestingly, the sonic flow coefficient slightly increases with the position along the high-porosity plates. It seems that the vena contracta area increases along the plate.

Remarkably, all the points for the first holes lie on one curve. This demonstrates that the first hole is mainly influenced by the diameter, even though an upstream influence of the following holes is observable. In contrast to Fig.~\ref{fig:CaneD}, the points for the last holes do not converge because the bleed mass flux, and hence the thinning of the boundary layer along the plate, are not similar.

\begin{figure}[b!]
    \centering
    \includegraphics[trim={0.cm 0.2cm 0.cm 0.2cm},clip]{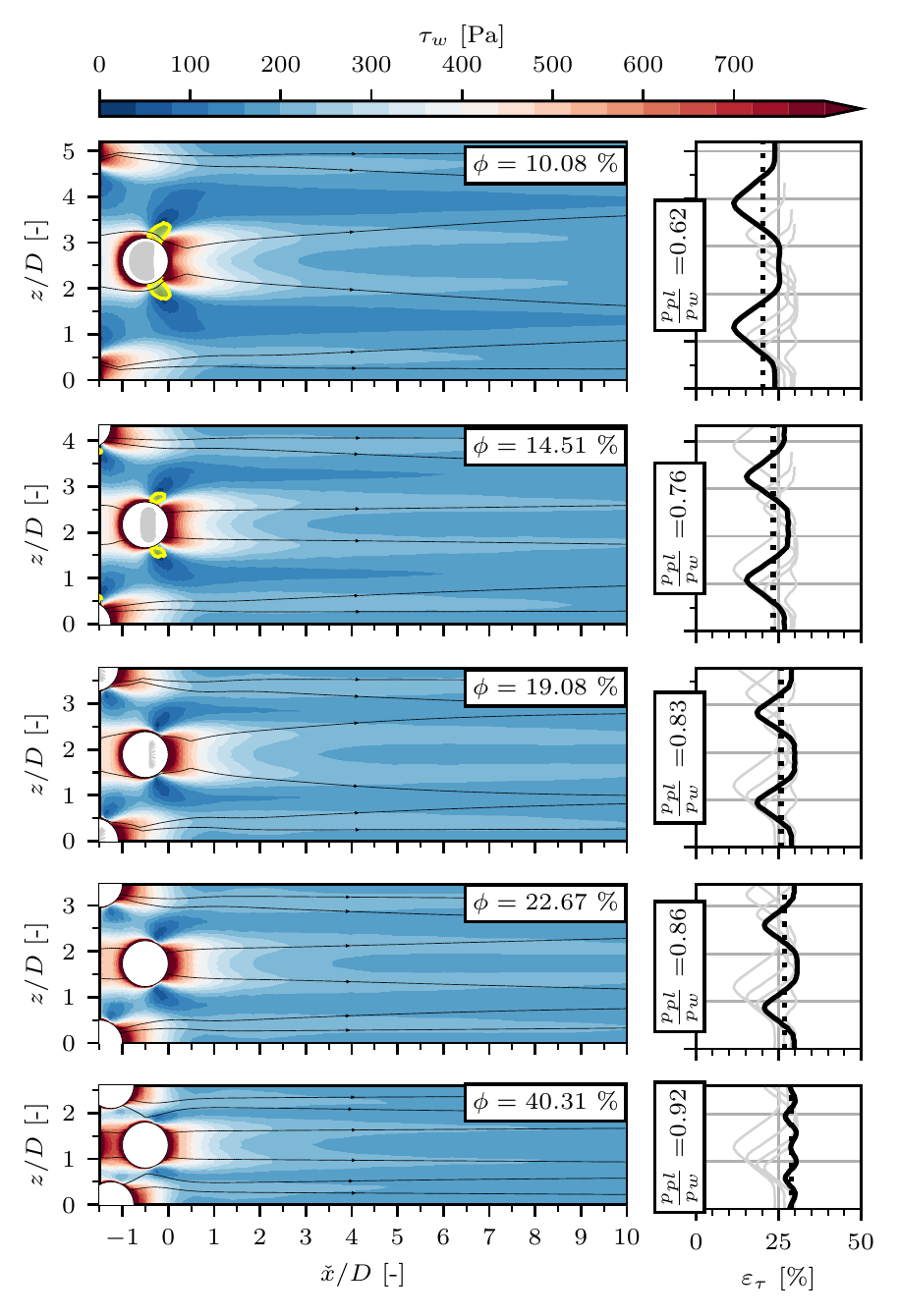}

    \begin{subfigure}[t!]{0.39\textwidth}
        \centering
        \caption{Nominal wall shear stress}
        \label{fig:shearPhinominal}
    \end{subfigure}
    \begin{subfigure}[t!]{0.25\textwidth}
        \centering
        \caption{Rise in wall shear stress}
        \label{fig:shearPhiincrease}
    \end{subfigure}
    \caption{Comparison of the wall shear stress downstream of the bleed region for different porosities; yellow areas in (a) visualize regions with negative streamwise component and gray areas indicate supersonic hole entry flow; (b) wall shear stress \SI{10}{\mm} ($20 D$) downstream of the last hole compared to its value \SI{10}{\mm} upstream of the first hole; dotted line illustrates averaged value along the spanwise direction}
    \label{fig:shearPhi}
\end{figure}

The influence of the porosity level on the wall shear stress for a constant mass flow rate is detailed in Fig.~\ref{fig:shearPhi}. Similar to the plate length, porosity has an essential impact on the removable mass flow rate. The lower the porosity, the lower the number of holes. Therefore, lower pressure ratios are required to obtain the same relative bleed mass flow rate as in the case of high porosity levels. Fig.~\ref{fig:shearPhi} shows the behavior for removing \SI{53}{\%} of the mass flow rate computed on the displacement thickness.

The general trend for the effect of the porosity is the following: the higher the porosity, the higher the effectiveness. The smaller spanwise distance is the key parameter, as the wall shear stress between the holes is lower. Thus, a larger distance leads to a less homogeneous flow downstream of the porous plate resulting in a lower average rise in the wall shear stress. The wall shear stress downstream on the hole-cutting plane is unaffected and similar for different porosity levels. Only for very low porosity a lower shear stress rise is observed as the intensity of the barrier shock is significantly higher.

As stated in the previous subsection, a lower rise in the wall shear stress corresponds to boundary layers with a higher shape factor. Thus, a more homogeneous boundary layer along the span is present in case of high porosities. On the contrary, lower porosities result to higher variations in the spanwise directions. Thus, fuller profiles are found downstream of the bleed holes, while the flow between the holes is less affected by the suction.

\begin{figure}[htb!]
    \centering
    \includegraphics[trim={0.cm 0.2cm 0.cm 0.2cm},clip]{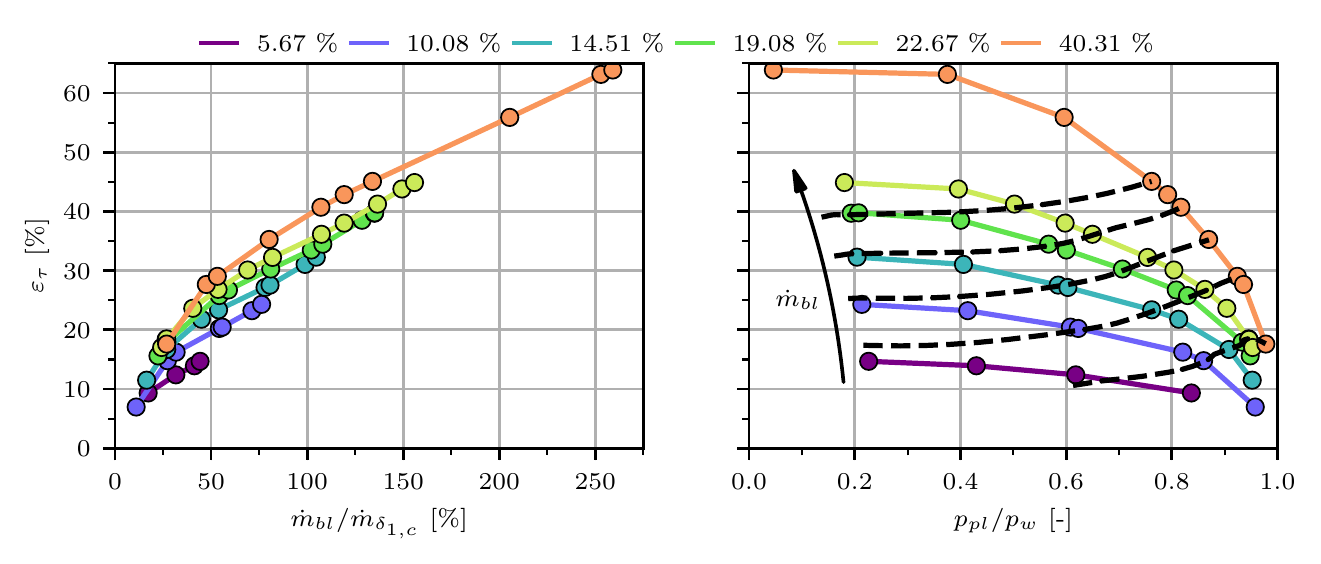}
    \begin{subfigure}[t!]{0.45\textwidth}
        \centering
        \caption{Rise in wall shear stress as function of the bleed mass flow rate}
        \label{fig:effectivenessPhim}
    \end{subfigure}
    \begin{subfigure}[t!]{0.42\textwidth}
        \centering
        \caption{Rise in wall shear stress as function of the pressure ratio}
        \label{fig:effectivenessPhip}
    \end{subfigure}
    \caption{Rise in the wall shear stress as function of the bleed mass flow rate $\dot{m}_{bl}$ and the pressure ratio $p_{pl}/p_w$ for varying porosities; dashed lines in (b) pinpoint lines of constant bleed mass flow rate}
    \label{fig:effectivenessPhi}
\end{figure}

The evaluation for different bleed mass flow rates is visualized in Fig.~\ref{fig:effectivenessPhi}. As already stated, a large porosity increases the maximal removable mass flow rate. Fig.~\ref{fig:effectivenessPhim} demonstrates this effect. A higher effectiveness for large porosities is revealed in Fig.~\ref{fig:effectivenessPhip}, where the dashed line illustrates constant bleed mass flow rates. Similar to the graph for the plate length, the best effect is achieved for pressure ratios $p_{pl}/p_w \approx 0.9$. For choking conditions, the thinning of the boundary layer is lower.

\subsection{Thickness-to-diameter ratio} \label{sec:TD}

The influence of the thickness-to-diameter ratio is considered in several existing bleed models~\cite{Harloff1996, Grzelak2021}. However, the flow topology that causes the change in the bleed efficiency is not sufficiently regarded. Therefore, the thickness-to-diameter ratio is varied from \SIrange{0.5}{8}{} in this study. In contrast to the previous subsections, the hole diameter is increased to \SI{1.0}{\mm} to reduce the mesh size and computational costs.

\begin{figure}[h!]
    \centering
    \includegraphics[trim={0.cm 0.2cm 0.cm 0.2cm},clip]{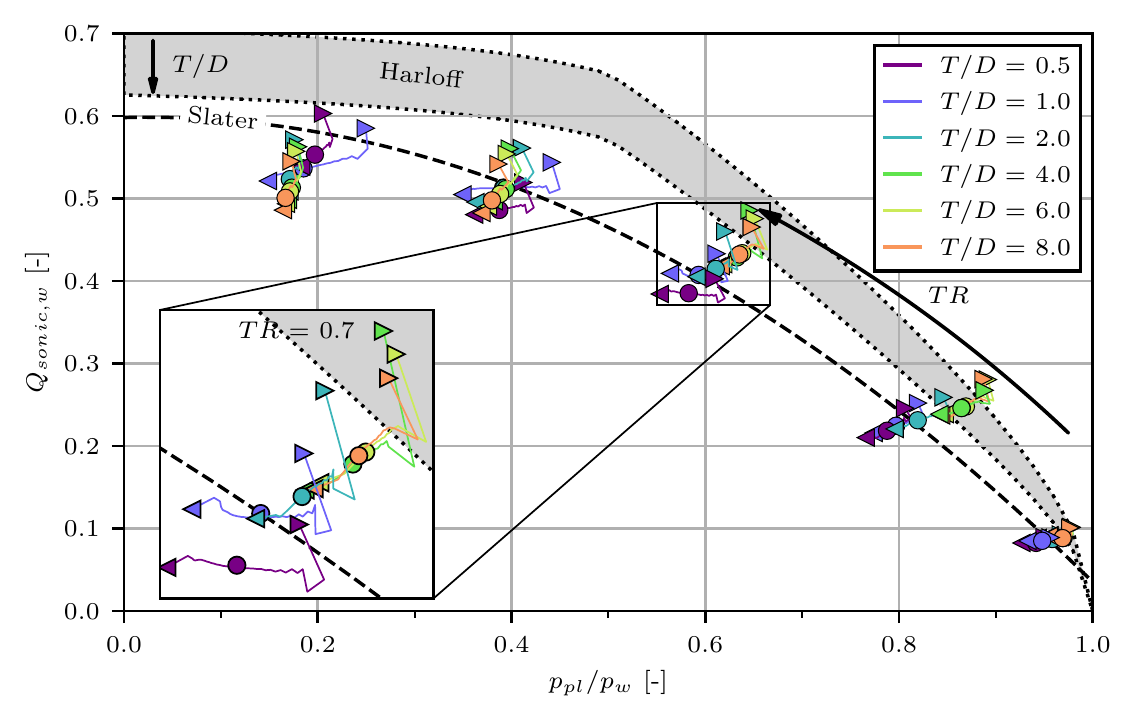}
    \caption{Surface sonic flow coefficient for different thickness-to-diameter ratios}
    \label{fig:CaneTD}
\end{figure}

The influence of the thickness-to-diameter ratio on the sonic flow coefficient is detailed in Fig.~\ref{fig:CaneTD}. Significant differences are apparent depending on the pressure ratio. For high pressure ratios, higher sonic flow coefficients and pressure ratios are found for high thickness-to-diameter ratios. A very low $T/D$ leads to major degradation of the efficiency. However, the differences are minor for large ratios with a maximum around $T/D=6$.

Interestingly, the trend is inverted for choking conditions, where higher sonic flow coefficients are obtained for small $T/D$. For $p_{pl}/p_w \approx 0.2$, the sonic flow coefficient for the thinnest plate is approximately \SI{10}{\%} higher than for the thickest plate. On the contrary, for $p_{pl}/p_w \approx 0.6$, the sonic flow coefficient for the thinnest plate is more than \SI{10}{\%} lower compared to the thickest plates.

\begin{figure}[htb!]
    \centering
    \includegraphics[trim={0mm 110mm 0mm 0mm},clip]{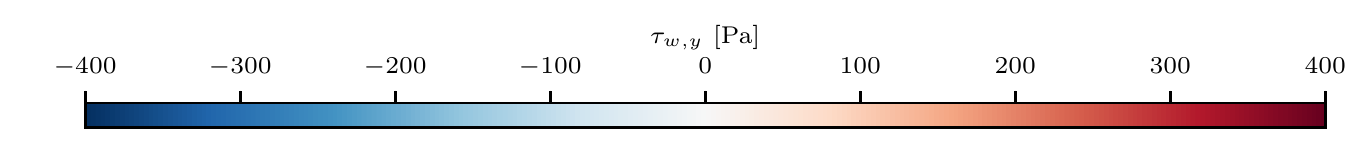}
    \begin{subfigure}[t!]{0.3\textwidth}
        \centering
        \includegraphics[trim={9mm 40mm 6mm 24mm},clip, width=\linewidth]{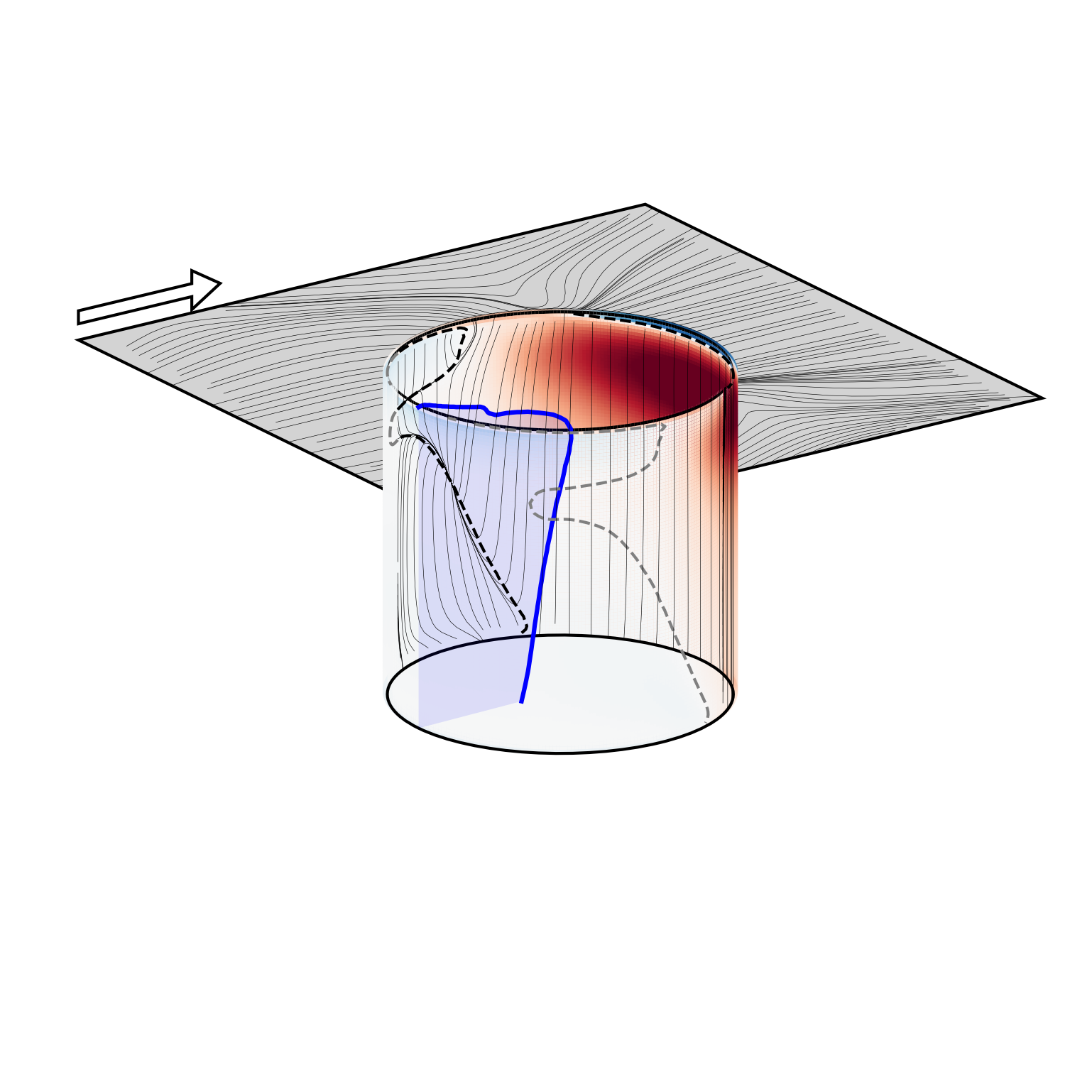}
        \caption{$T/D=1$; $TR=0.3$}
        \label{fig:holeShearA}
    \end{subfigure}
    \begin{subfigure}[t!]{0.3\textwidth}
        \centering
        \includegraphics[trim={9mm 40mm 6mm 24mm},clip, width=\linewidth]{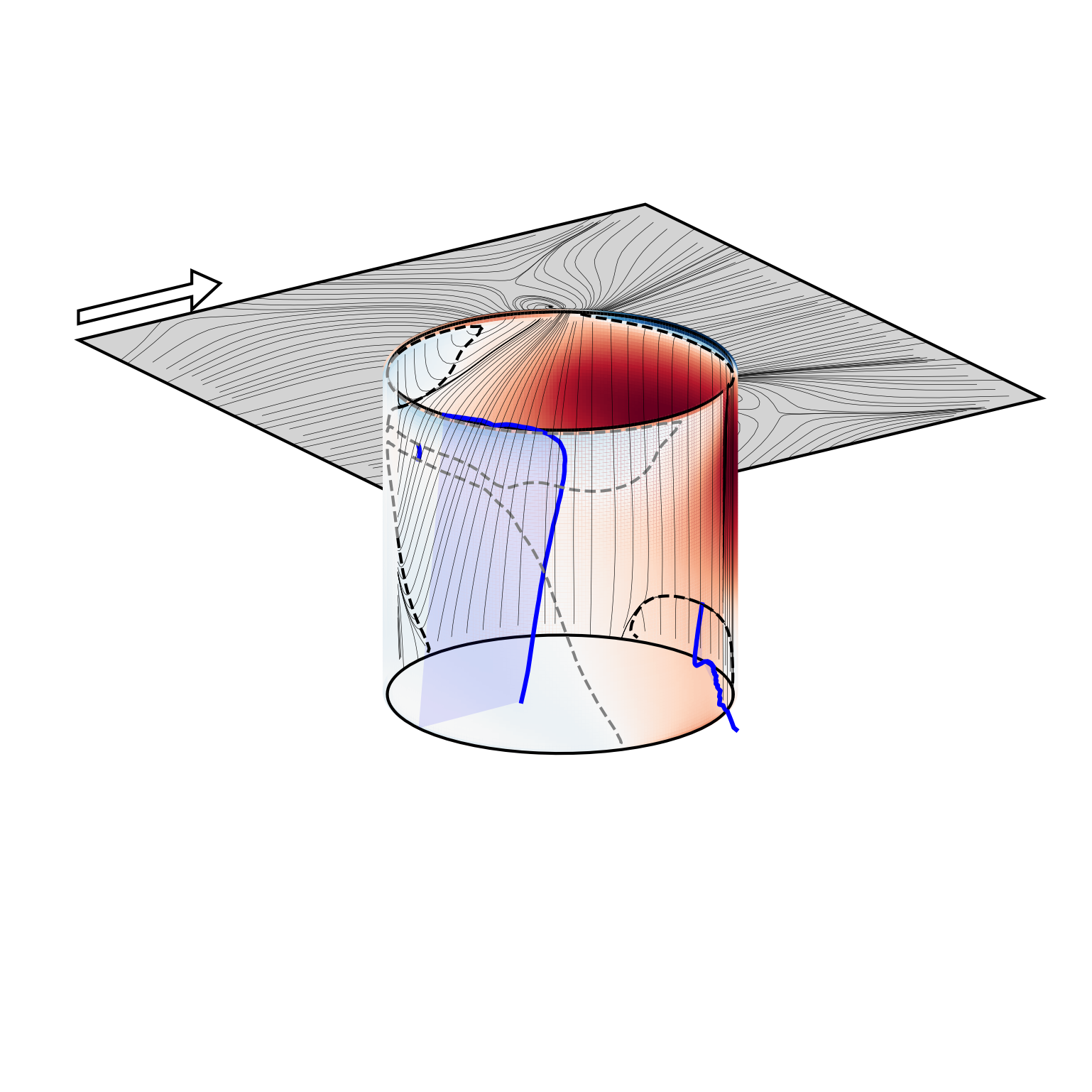}
        \caption{$T/D=1$; $TR=0.7$}
        \label{fig:holeShearB}
    \end{subfigure}
    \begin{subfigure}[t!]{0.3\textwidth}
        \centering
        \includegraphics[trim={9mm 40mm 6mm 24mm},clip, width=\linewidth]{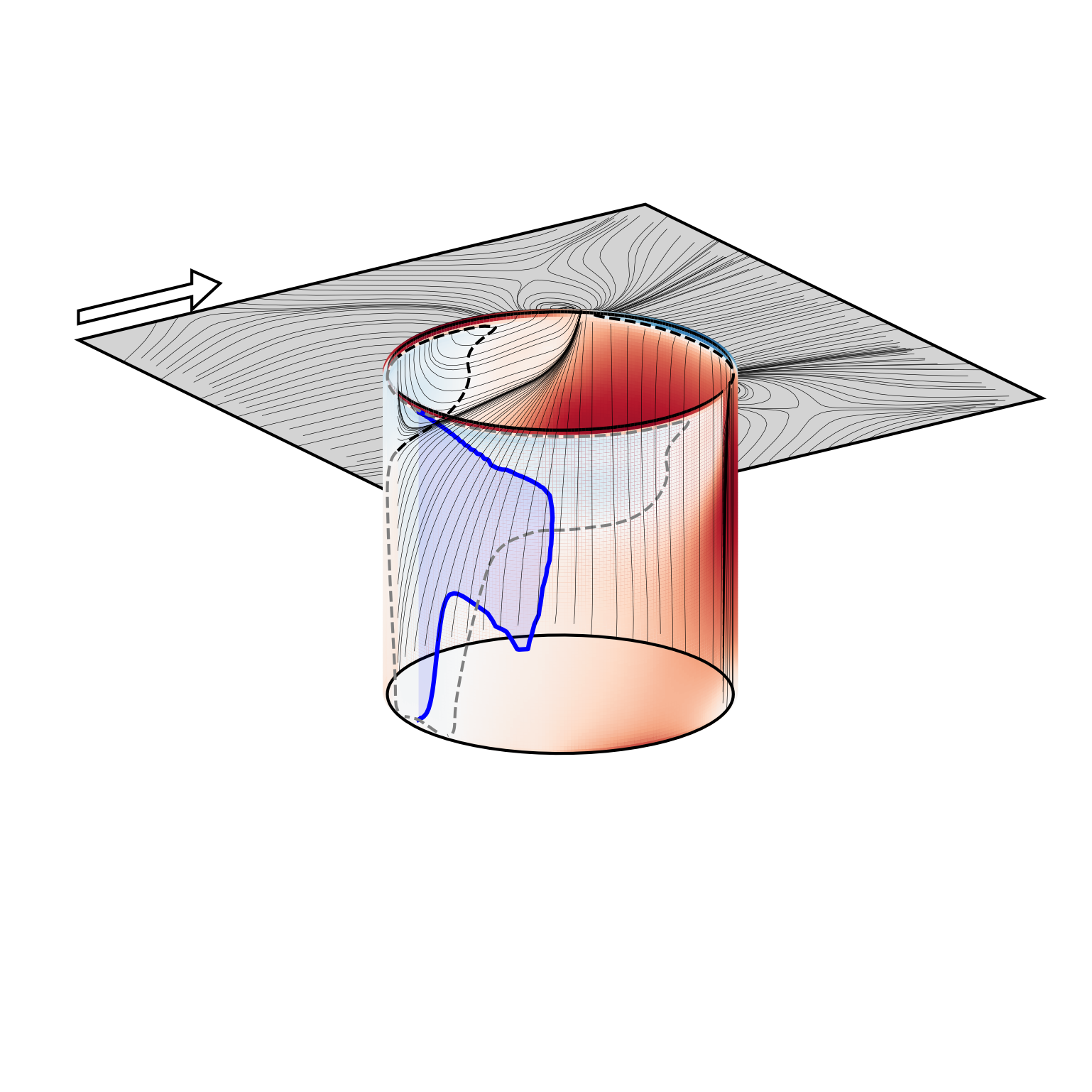}
        \caption{$T/D=1$; $TR=1.3$}
        \label{fig:holeShearC}
    \end{subfigure}

    \begin{subfigure}[t!]{0.3\textwidth}
        \centering
        \includegraphics[trim={17mm 27mm 14mm 12mm},clip, width=\linewidth]{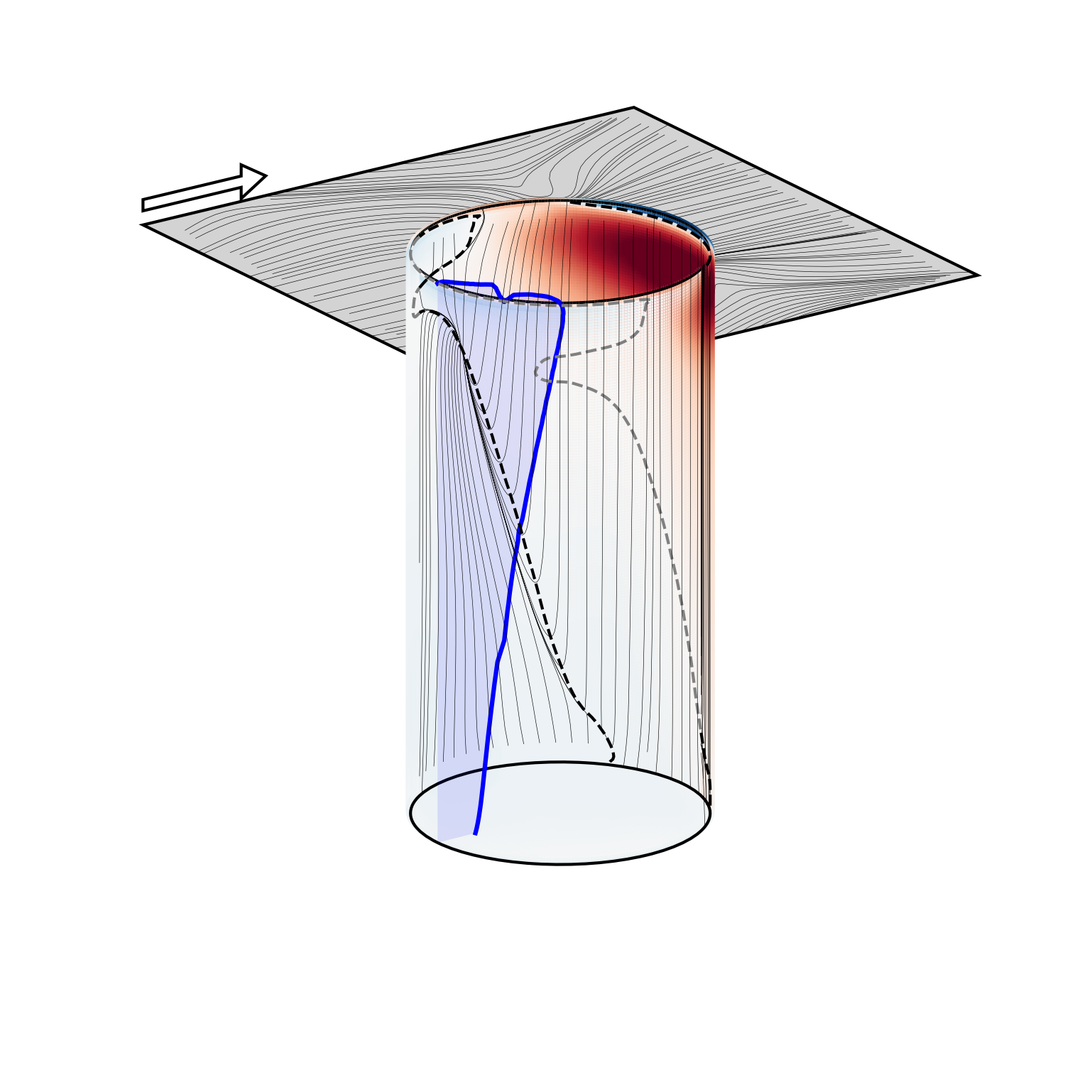}
        \caption{$T/D=2$; $TR=0.3$}
    \end{subfigure}
    \begin{subfigure}[t!]{0.3\textwidth}
        \centering
        \includegraphics[trim={17mm 27mm 14mm 12mm},clip, width=\linewidth]{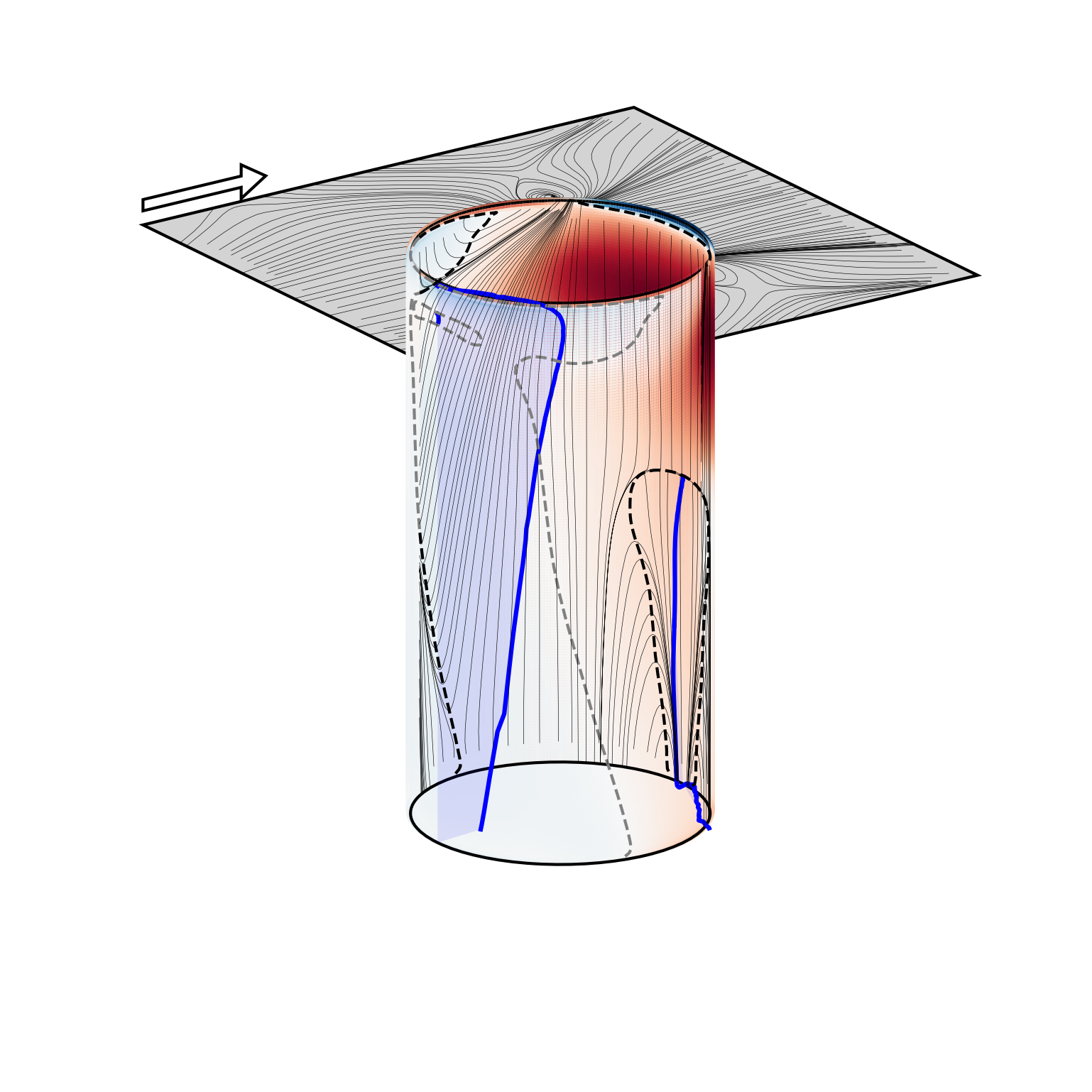}
        \caption{$T/D=2$; $TR=0.7$}
    \end{subfigure}
    \begin{subfigure}[t!]{0.3\textwidth}
        \centering
        \includegraphics[trim={17mm 27mm 14mm 12mm},clip, width=\linewidth]{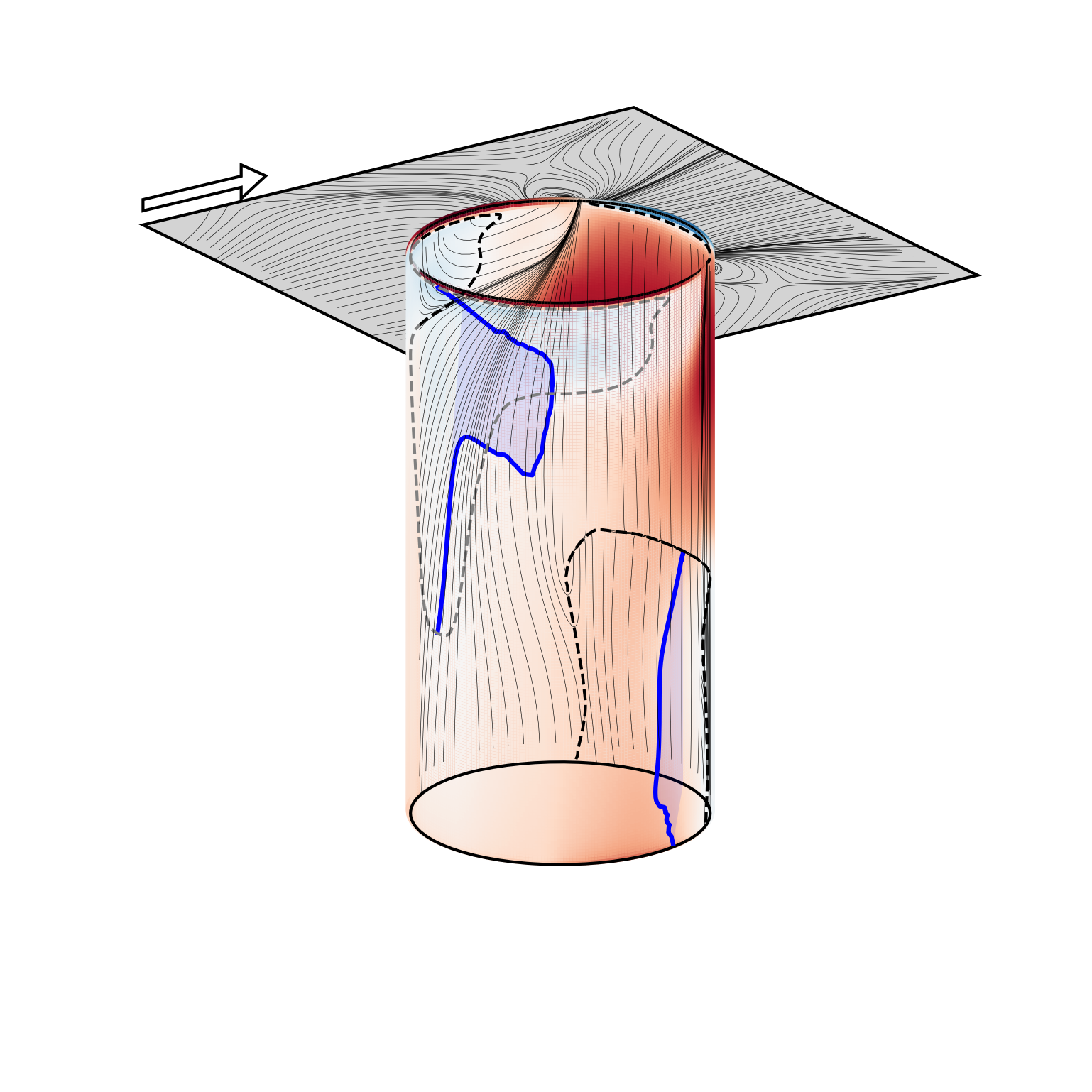}
        \caption{$T/D=2$; $TR=1.3$}
    \end{subfigure}

    \begin{subfigure}[t!]{0.3\textwidth}
        \centering
        \includegraphics[trim={31mm 14.7mm 28mm 3mm},clip, width=\linewidth]{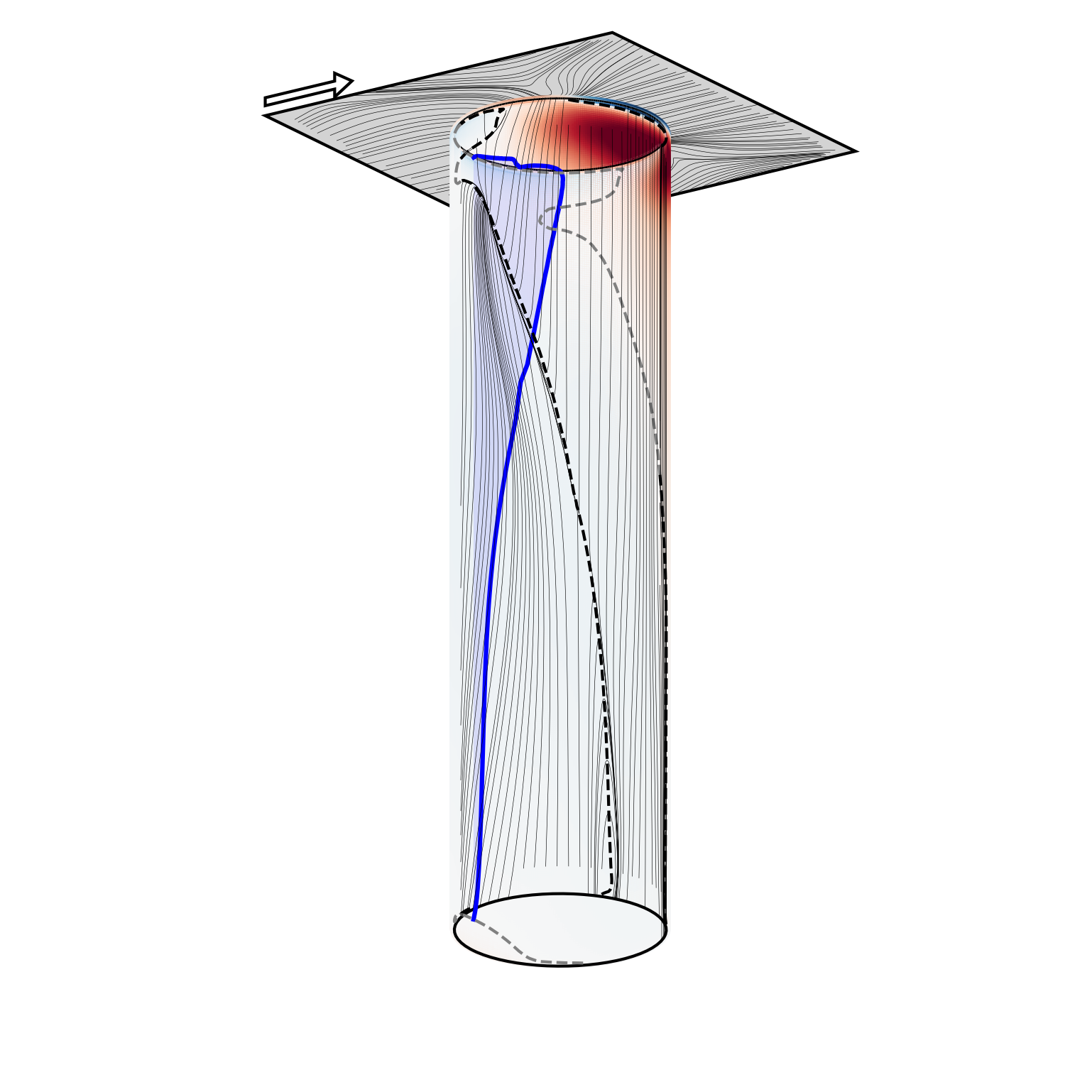}
        \caption{$T/D=4$; $TR=0.3$}
    \end{subfigure}
    \begin{subfigure}[t!]{0.3\textwidth}
        \centering
        \includegraphics[trim={31mm 14.7mm 28mm 3mm},clip, width=\linewidth]{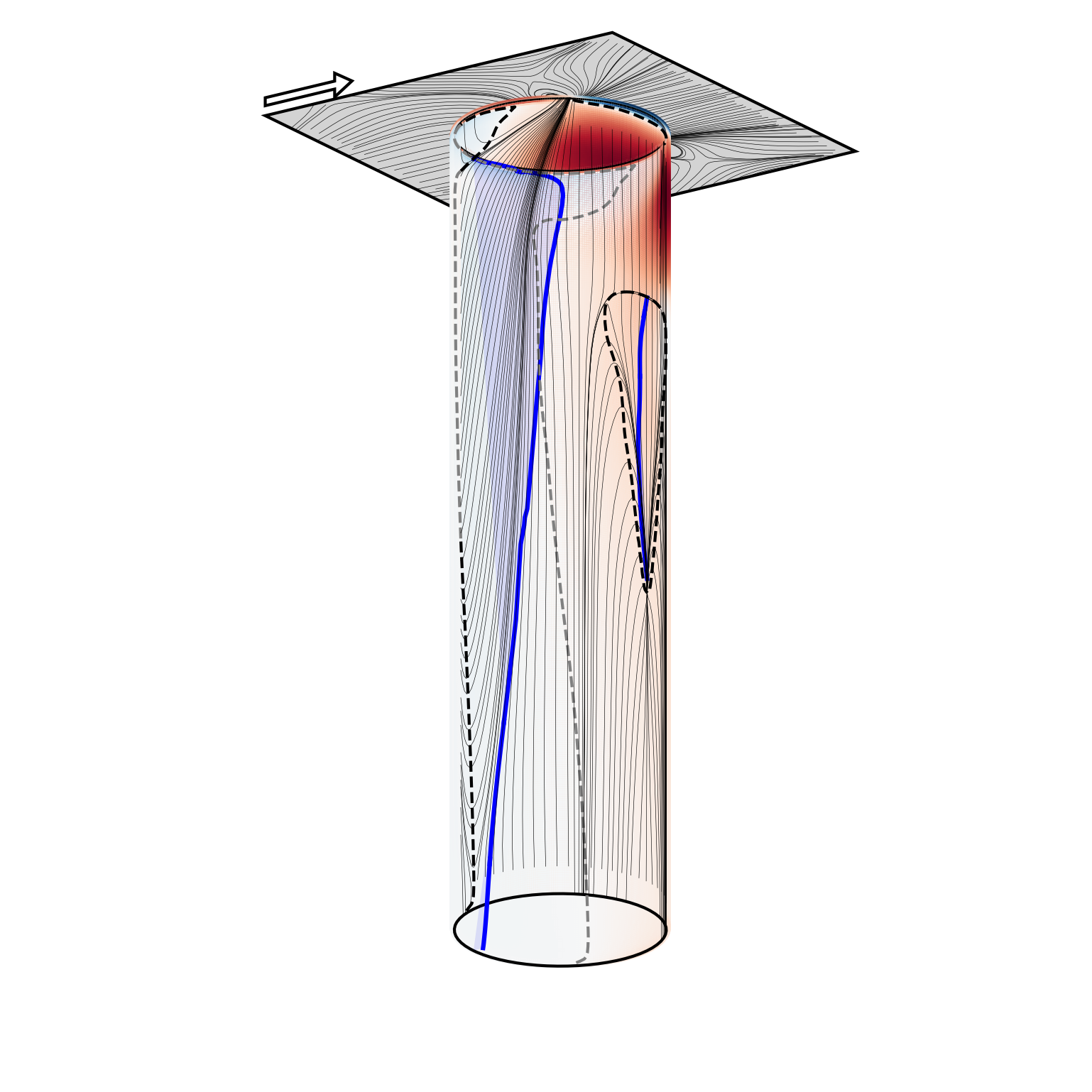}
        \caption{$T/D=4$; $TR=0.7$}
    \end{subfigure}
    \begin{subfigure}[t!]{0.3\textwidth}
        \centering
        \includegraphics[trim={31mm 14.7mm 28mm 3mm},clip, width=\linewidth]{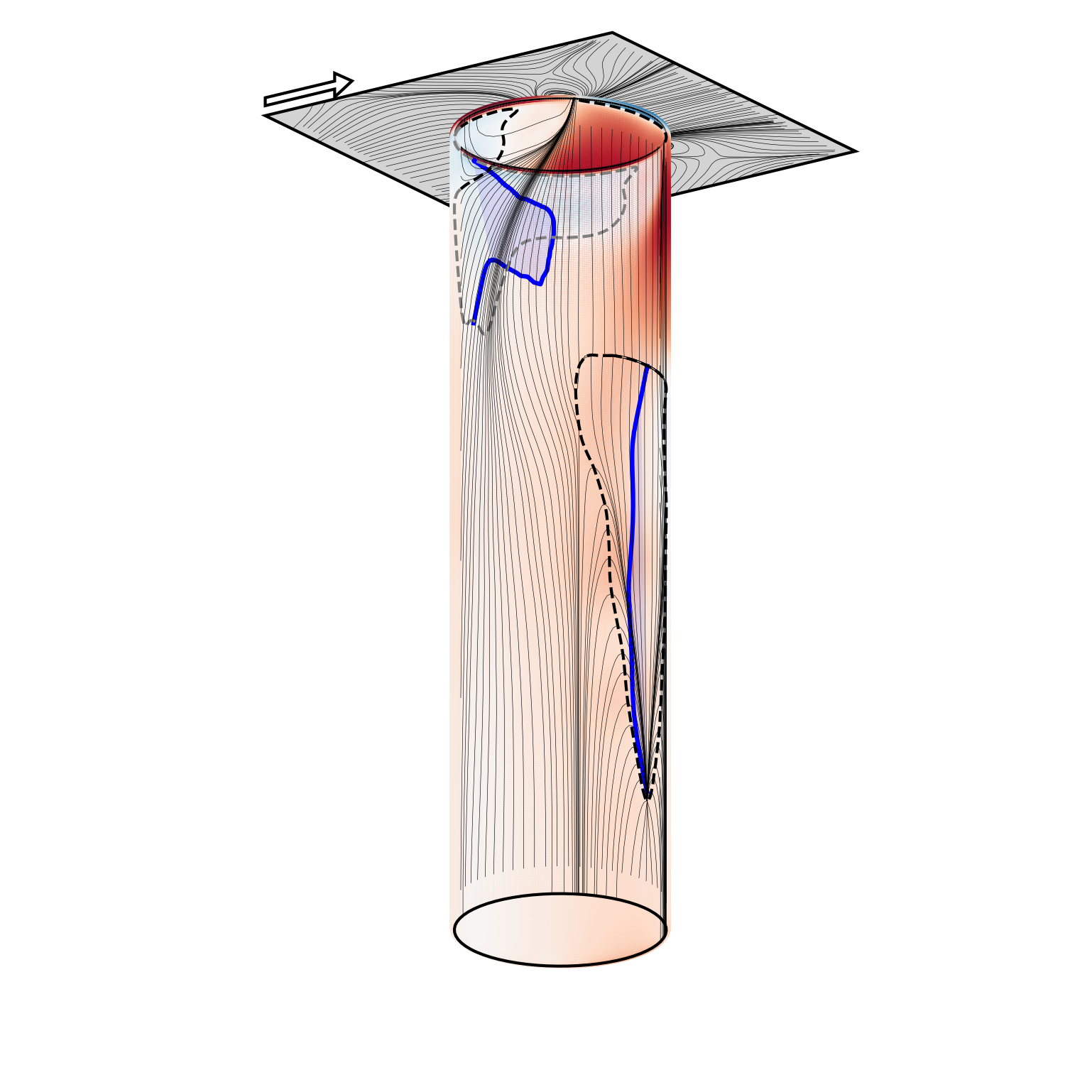}
        \caption{$T/D=4$; $TR=1.3$}
    \end{subfigure}
    \caption{Wall shear stress inside the holes at $\hat{x}=$ \SI{8}{\mm} ($16D$); dashed lines illustrate $\tau_{w,y}=$ \SI{0}{\Pa} isolines; blue patches visualize areas with upward-directed momentum on the hole-cutting plane}
    \label{fig:holeShear}
    \begin{tikzpicture}[remember picture, overlay]
        \draw [black, -{Stealth[length=1.5mm, width=1.5mm]}, line width=0.25mm] (6.35, 9.80) -- (5.40, 10.50);
        \node[text centered, text width=0.5cm] at (6.65, 9.80) {($\bm{\ast}$)};
        \draw [black,-{Stealth[length=1.5mm, width=1.5mm]}, line width=0.25mm] (-6.35, 14.80) -- (-5.50, 14.60);
        \node[text centered, text width=0.5cm] at (-7.10, 14.80) {\shortstack{\small Hole-\\\small cutting\\\small plane}};
    \end{tikzpicture}
\end{figure}
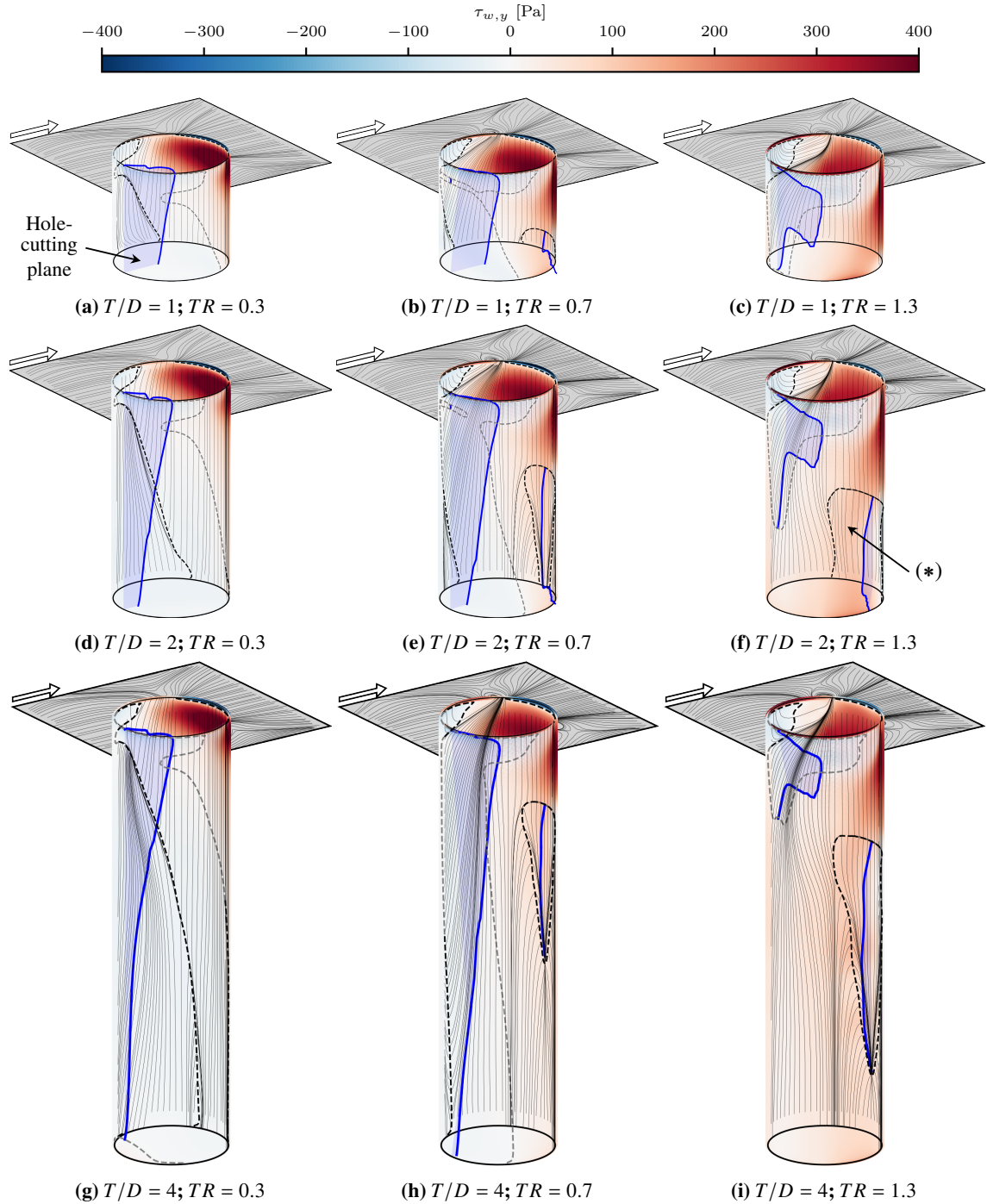

Moreover, significant differences in the pressure ratios are apparent for unchoked conditions (low $TR$). This results from higher pressure losses caused by the flow inside the holes. In order to explain it, the wall shear stress and the friction lines inside the holes are visualized in Fig.~\ref{fig:holeShear} for $T/D=$ \SIlist{1; 2; 4}{} (top to bottom) and $TR=$ \SIlist{0.3; 0.7; 1.3}{} (left to right). The holes are shown from the bottom-up view, and the external flow goes from the left to the right. The gray patches represent the external wall with the friction lines around the holes illustrated. Inside the holes, the wall shear stress contours in the vertical direction (the main direction of the transpiration flow) are presented on the wall. The dashed lines present the $\tau_{w,y}=0$ isolines to highlight the areas of reversed flow, the solid lines the friction lines on the anterior face. Moreover, the separated areas on the hole-cutting plane are illustrated by the blue patches.

The first column of Fig.~\ref{fig:holeShear} shows the flow inside the holes for large pressure ratios where high throat-to-diameter ratios are beneficial. Inside the holes, a large separated area is observed independently of the plate thickness. However, its size on the hole-cutting plane decreases with further distance to the hole entry. For a ratio $T/D=4$, the separated region is almost closed; thus, only a small amount of flow streams from the cavity inside the hole. In contrast, a significant fraction of the outlet cross section for thin plates is covered by reversed flow streaming into the hole. It is assumed that higher pressure losses are a consequence and lead to lower pressure ratios compared to higher thickness-to-diameter ratios, as shown in Fig.~\ref{fig:CaneTD}. Moreover, all mass sucked from the cavity into the hole must also exit the hole again. As a result, the bleed mass flow rate is assumed to be lower, as indicated by the sonic flow coefficient.

In the second column, similar effects are apparent even though the pressure ratio is lower as the throat ratio increases. The shape of the separated region on the hole-cutting plane is equal for all plate thicknesses. However, a strong bending of the friction lines close to the hole entry towards the front of the hole is notable. Consequently, the wall areas with negative shear stress decrease in size compared to those with larger pressure ratios. Thus, less air streams from the cavity into the hole independently of the thickness, which explains the lower differences between the thickness-to-diameter ratios, as shown in Fig.~\ref{fig:CaneTD}.

The last column illustrates the throat ratio $TR=1.3$, corresponding to $p_{pl}/p_w \approx 0.4$. According to Fig.~\ref{fig:CaneTD}, a ratio of $T/D=1$ results in the highest sonic flow coefficient. Contrary to higher pressure ratios, the separated region in the front of the holes is closed, even for the thinnest plate. Thus, there is only a recirculation of flow inside the hole but no suction from the cavity. As a result, the sonic flow coefficient is higher. For $T/D=2$, the sonic flow coefficient is the lowest.

In contrast to the thinnest plate, a second separated area at the rear of the hole is present ($\bm{\ast}$), which causes a flow from the cavity into the hole. Also, a second separated area is apparent for the thickest plate. However, it is closed, and the flow is fully attached at the hole outlet leading to an increase in the sonic flow coefficient. Nevertheless, the pressure ratio is slightly lower than for $T/D=1$, resulting from friction drag losses inside the hole because of the longer length of the attached flow.

Our findings do not support the assumption of~\citet{Harloff1996} that ratios $T/D > 3$ lead to a degradation of the sonic flow coefficient, as shown in Fig.~\ref{fig:CaneTD}. In contrast, higher thickness-to-diameter ratios improve the performance of the porous bleed for unchoked conditions. The reason is the large separated region at the front of the hole caused by the external flow. For low suction rates, its size is maximal, and it may lead to inflow from the cavity caused by the entrainment of the jet inside the hole. With a longer hole, the flow is fully attached, and inflow from the cavity is prevented.

Interestingly, the thickness-to-diameter ratio has no significant effect on the external flow. As apparent in Fig.~\ref{fig:holeShear}, the friction lines on the external wall are almost identical, independent of the plate thickness. Also, only marginal differences in the wall shear stress are notable downstream of the porous bleed region. Thus, the conclusion of no significant influence of the thickness-to-diameter ratio on the effectiveness of the porous bleed is made.


\subsection{Stagger angle} \label{sec:beta}

The last investigated parameter is the stagger angle which defines the hole pattern on the plate. In this study, the angle takes values of \SIlist{30; 45; 60; 90}{\degree}. The plates with $\beta=$ \SI{30}{\degree} and $\beta=$ \SI{60}{\degree} are equal with the holes arranged in the shape of an equilateral triangle but rotated by \SI{90}{\degree}. For the plates with $\beta=$ \SIlist{45; 90}{\degree}, the holes are arranged in a square shape. Thus, every second column of holes is located at the same streamwise position.

\begin{figure}[htb!]
    \centering
    \includegraphics[trim={0.cm 0.2cm 0.cm 0.2cm},clip]{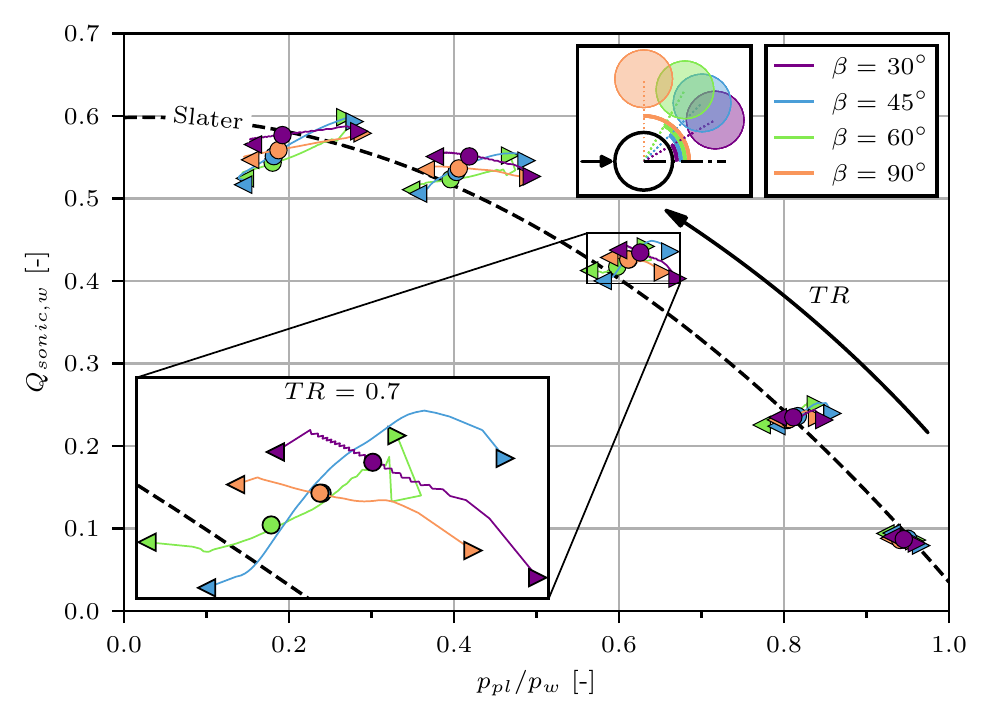}
    \caption{Surface sonic flow coefficient for different stagger angles}
    \label{fig:CaneBeta}
\end{figure}

\begin{figure}[b!]
    \centering
    \includegraphics[trim={0.cm 0.2cm 0.cm 0.2cm},clip]{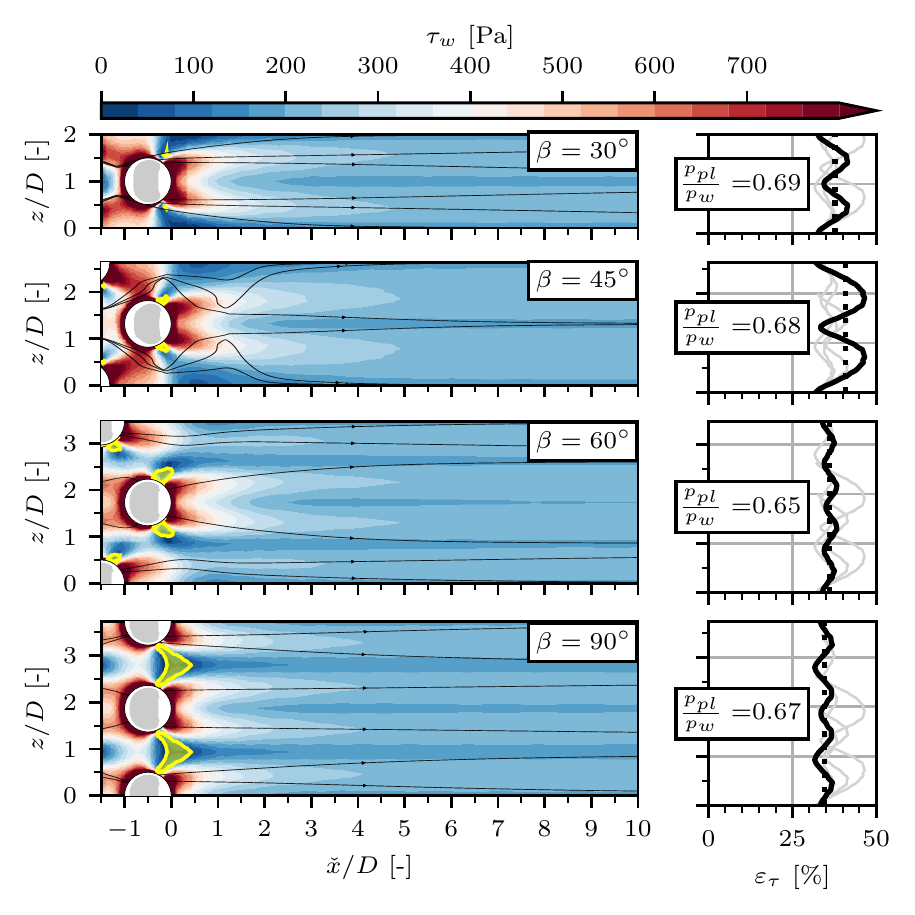}

    \begin{subfigure}[t!]{0.39\textwidth}
        \centering
        \caption{Nominal wall shear stress}
    \end{subfigure}
    \begin{subfigure}[t!]{0.25\textwidth}
        \centering
        \caption{Rise in wall shear stress}
    \end{subfigure}
    \caption{Comparison of the wall shear stress downstream of the bleed region for different stagger angles; yellow areas in (a) visualize regions with negative streamwise component; (b) wall shear stress \SI{10}{\mm} ($20 D$) downstream of the last hole compared to its value \SI{10}{\mm} upstream of the first hole; dotted line illustrates averaged value along the spanwise direction}
    \label{fig:shearBeta}
\end{figure}

The surface sonic flow coefficient for the different stagger angles is shown in Fig.~\ref{fig:CaneBeta}. Independently of the pressure ratio, the mean sonic flow coefficient is the highest for $\beta=$ \SI{30}{\degree} and the lowest for \SI{60}{\degree} even so the plates are identical but only rotated. This results from the longer streamwise distance between the holes that induces a lower momentum upstream of the hole as the wall area where the boundary layer thickens is longer. However, the opposite trend is found for the first holes. There, the spanwise distance is assumed to be more relevant, which is in line with the findings from Sec.~\ref{sec:Eichorn}, where the degradation of the sonic flow coefficient is stated for multi-column plates.

The wall shear stress downstream of the porous plate for the different stagger angles is visualized in Fig.~\ref{fig:shearBeta}. The removed bleed mass flow rate corresponds to \SI{107}{\%} of the inflow displacement mass flow rate. Similar to the sonic flow coefficient, the variation of the span- and streamwise distances between the holes lead to a variation in the effectiveness. The smaller the spanwise distance, the lower the size of the area with the reversed flow. This leads to a more significant rise in wall shear stress.

However, with lower spanwise distances between the holes, also the intensity of the barrier shock increases, which leads, in turn to a decrease in the wall shear stress on the hole-cutting plane downstream of the holes. Consequently, the flow field downstream of the plate is less homogeneous. For $\beta=$ \SI{60}{\degree}, the highest rise in the wall shear stress is found by having the most extensive variations along the spanwise direction. The plates with a stagger angle of \SIlist{60; 90}{\degree} affect the flow more homogeneously but less drastically.


\subsection{Comparison of all parameters}

\begin{figure}[b!]
    \centering
    \includegraphics[trim={0.cm 0.2cm 0.cm 0.2cm},clip]{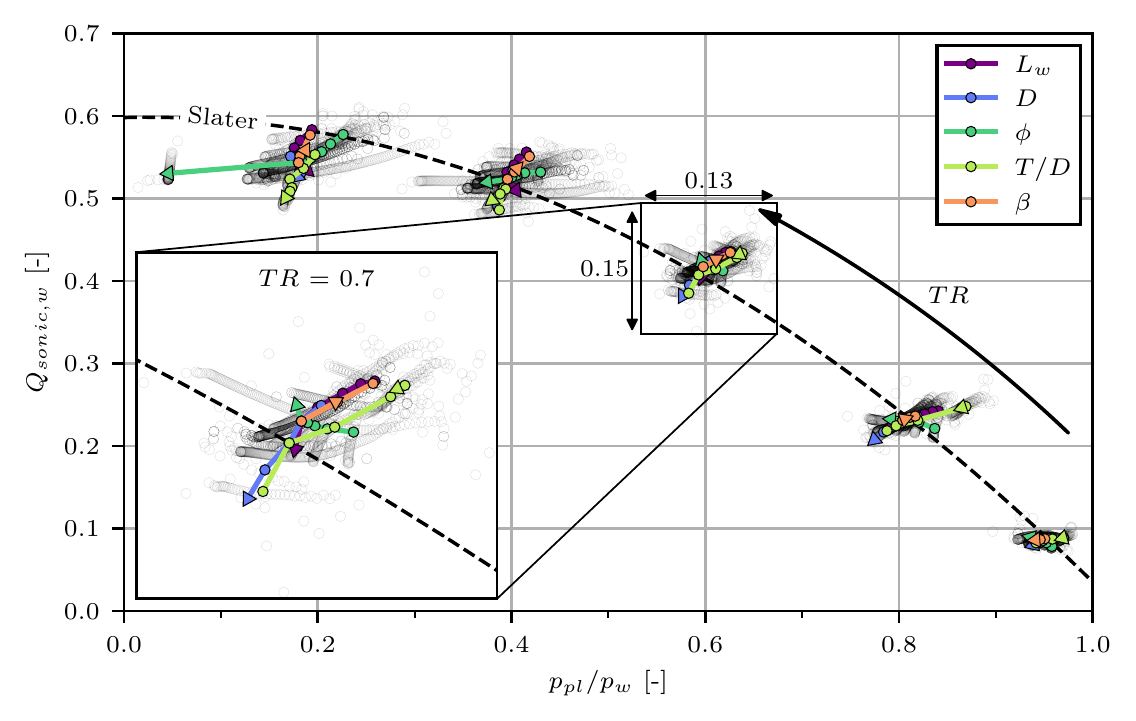}
    \caption{Surface sonic flow coefficient for all parameters}
    \label{fig:CaneAll}
\end{figure}

The influence of the investigated parameters on the sonic flow coefficient are gathered in Fig.~\ref{fig:CaneAll}. The colored lines show the global trend of the plates for a variation of one parameter, and the gray circles represent the spreading of all locally extracted sonic flow coefficients. The arrows show the direction in which the parameter grows. Therefore, it must be noted that the thickness-to-diameter ratio and the stagger angle show no steady trend but have an optimum. As a result, the highest value is not equal to the maximum or minimum.

\begin{figure}[t!]
    \centering
    \includegraphics[trim={0.cm 0.2cm 0.cm 0.2cm},clip]{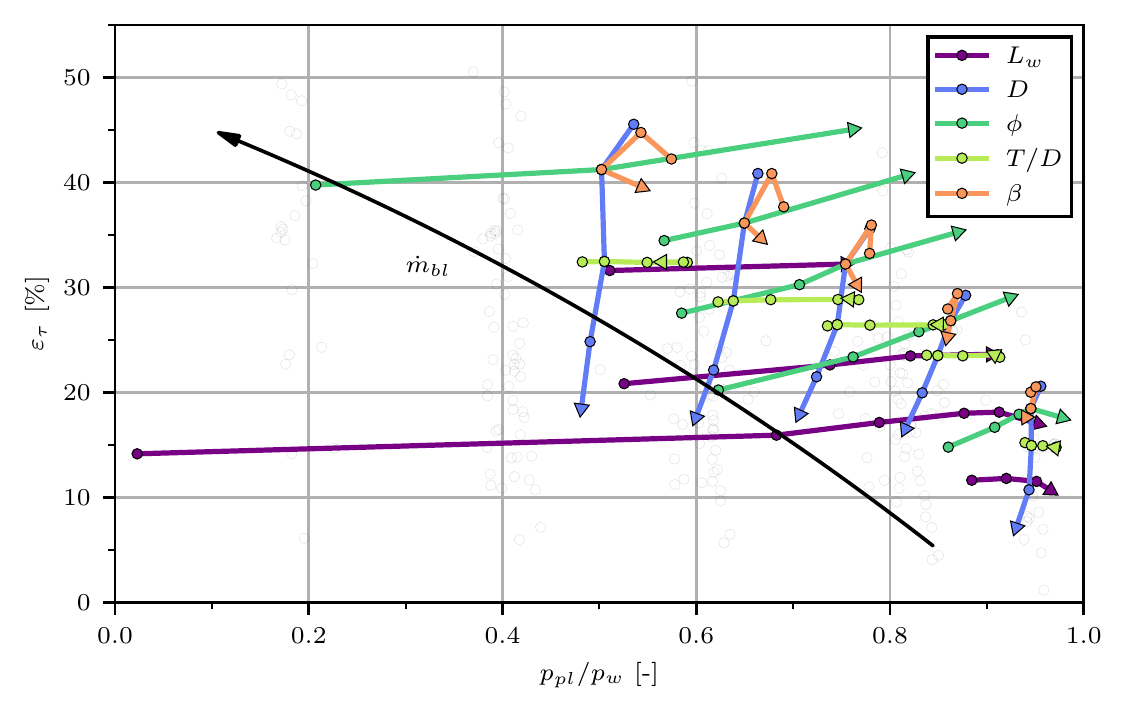}
    \caption{Rise in the wall shear stress for all parameters}
    \label{fig:ShearAll}
\end{figure}

The global trends are generally similar: an increase in the plate length or hole diameter leads to lower sonic flow coefficients and pressure ratios. Thus, lower values are more efficient. Also, the pressure ratio decreases for higher porosities. However, the sonic flow coefficient is less affected and slightly increases for high pressure ratios. As stated in the previous section, the best stagger angle regarding the efficiency is $\beta=$ \SI{30}{\degree}, resulting in lower flow coefficients and plenum pressures for other values. High thickness-to-diameter ratios are preferable for high pressure ratios, while thin plates are more efficient for choked conditions.

While the global data is close to the regression of~\citet{Slater2012}, the local values can drastically differ from this trend. For a throat ratio of $TR=0.7$, the spreading of the sonic flow coefficient corresponds to more than \SI{35}{\%} of the estimated value. For other parameter combinations, the deviation might be higher.

The differences of the investigated porous plates are more significant if the effectiveness is regarded, as visualized in Fig.~\ref{fig:ShearAll}. Colored lines represent a parameter variation for a fixed bleed mass flow rate. A steep slope of a line highlights a strong parameter impact on the effectiveness. For example, large hole diameters result in a significantly lower thinning of the boundary layer and hence in a weaker rise in the wall shear stress. On the contrary, the length-to-diameter ratio does not affect the thinning of the boundary layer. However, the required plenum pressure to obtain the desired effect differs depending on the efficiency.

Altogether, the parameter can be ranked regarding effectiveness: the hole diameter has the most decisive impact, followed by porosity, stagger angle, plate length, and the thickness-to-diameter ratio having the lowest influence.

\section{Conclusion and future work}

This paper studied the influence of the plate length, hole diameter, porosity, thickness-to-diameter ratio, and the stagger angle on the performance of a porous bleed. The bleed efficiency and the effectiveness in thinning the turbulent boundary layer were presented. Efficiency in the form of the sonic flow coefficient and effectiveness as the rise in the wall shear stress have been evaluated as function of the pressure ratio from cavity plenum to external flow.

The findings of the study prove the influence of the hole diameter on both effectiveness and efficiency. The smaller the hole diameter, the better the thinning of the boundary layer. Moreover, small holes are more efficient as they enable the removal of higher mass flow rates for equal pressure ratios. The choice of porosity is, contrary to the hole diameter, a trade-off between effectiveness and efficiency. High-porosity plates guarantee the suction of sufficient mass flow rates to thin the boundary layer. However, high porosity levels correspond to an increased number of holes and, thus, to more pressure losses induced by the barrier shocks and expansion fans.

Plate length and thickness-to-diameter ratio mainly influence the efficiency as long, and thin plates lead to an increase in the pressure losses. The optimal thickness-to-diameter ratio is found when the flow at the hole exit is fully attached, and no air is sucked into the holes. However, the impact of the two parameters on the effectiveness is negligible, especially for the thickness-to-diameter ratio. The stagger angle influences both efficiency and effectiveness. Angles of $\beta=$ \SIrange{45}{60}{\degree} seem to be the optimum as the spanwise distance is not larger than the streamwise distance.

The present findings suggest several implications for existing porous bleed models, which do not consider the geometrical design of a porous plate. Moreover, the importance of the inflow conditions (Mach number and boundary layer profile) needs to be investigated.

\section*{Acknowledgments}

This project has received funding from the European Union’s Horizon 2020 research and innovation programme under grant agreement No EC grant 860909.

\end{document}